\documentclass[12pt,journal,onecolumn]{IEEEtran}
\IEEEoverridecommandlockouts
\usepackage{bm}
\usepackage{cite}
\usepackage[cmex10]{amsmath}
\usepackage{tabularx,colortbl}
\definecolor{Gray}{gray}{0.9}
\usepackage{amsthm,mathrsfs,dsfont}
\usepackage{amsmath}
\usepackage{mathtools}
\usepackage{amssymb}
\usepackage{float}
\usepackage{booktabs}
\usepackage{threeparttable}
\usepackage{multirow}
\usepackage{subfigure}
\usepackage{stfloats}
\usepackage{subeqnarray}
\usepackage[linesnumbered, ruled, lined]{algorithm2e}
\SetKwRepeat{Do}{do}{while}
\allowdisplaybreaks[4]
\DeclareMathOperator*{\argmax}{argmax}

\newtheorem{theorem}{Theorem}
\newtheorem{lemma}{Lemma}

\ifCLASSINFOpdf
  \usepackage[pdftex]{graphicx}
  \DeclareGraphicsExtensions{.pdf,.jpeg,.png}
\else
  \usepackage[dvips]{graphicx}
  \DeclareGraphicsExtensions{.eps}
\fi

\begin{document}

\title{Beam Allocation for Millimeter-Wave MIMO Tracking Systems}

\author{Deyou Zhang, Ang Li, He Chen, Ning Wei, Ming Ding, Yonghui Li, and \\ Branka Vucetic \vspace{-1em} \thanks{Deyou Zhang, Ang Li, He Chen, Yonghui Li, and Branka Vucetic are with the School of Electrical and Information Engineering, The University of Sydney, NSW 2006, Australia (email: \{deyou.zhang, ang.li2, he.chen, yonghui.li, branka.vucetic\}@sydney.edu.au). Ning Wei is with the National Key Laboratory of Science and Technology on Communications, University of Electronic Science and Technology of China, Chengdu, Sichuan, China (email: wn@uestc.edu.cn). Ming Ding is with Data61, CSIRO, Sydney, NSW 2015, Australia (e-mail: Ming.Ding@data61.csiro.au). Part of this paper was presented at the IEEE International Conference on Communications, 2018 \cite{My-ICC}.}}

\maketitle

\begin{abstract}
In this paper, we propose a new beam allocation strategy aiming to maximize the average successful tracking probability (ASTP) of time-varying millimeter-wave MIMO systems. In contrast to most existing works that employ one transmitting-receiving (Tx-Rx) beam pair once only in each training period, we investigate a more general framework, where the Tx-Rx beam pairs are allowed to be used repeatedly to improve the received signal powers in specific directions. In the case of orthogonal Tx-Rx beam pairs, a power-based estimator is employed to track the time-varying AoA and AoD of the channel, and the resulting training beam pair sequence design problem is formulated as an integer nonlinear programming (I-NLP) problem. By dividing the feasible region into a set of subregions, the formulated I-NLP is decomposed into a series of concave sub I-NLPs, which can be solved by recursively invoking a nonlinear branch-and-bound algorithm. To reduce the computational cost, we relax the integer constraints of each sub I-NLP and obtain a low-complexity solution via solving the Karush-Kuhn-Tucker conditions of their relaxed problems. For the case when the Tx-Rx beam pairs are overlapped in the angular space, we estimate the updated AoA and AoD via an orthogonal matching pursuit (OMP) algorithm. Moreover, since no explicit expression for the ASTP exists for the OMP-based estimator, we derive a closed-form lower bound of the ASTP, based on which a favorable beam pair allocation strategy can be obtained. Numerical results demonstrate the superiority of the proposed beam allocation strategy over existing benchmarks.
\end{abstract}


\IEEEpeerreviewmaketitle

\section{Introduction}
In recent years, with the rapid penetration of mobile broadband Internet and multimedia applications, the ever-increasing data rate demand has made current sub-6GHz bands unprecedentedly crowded. Thanks to the rich available spectrum, millimeter-wave (mmWave) communication ranging from 30GHz to 300GHz emerged as a promising solution. However, due to the high frequency, mmWave signals suffer from severe propagation loss and atmospheric absorption \cite{MmWave-Survey, Relay-Probing}. Fortunately, the short operating wavelength enables a large antenna array to be integrated in a compact form, providing a considerable beamforming gain to compensate for the path loss in mmWave radio links \cite{Beamforming}. Nevertheless, such beamforming requires accurate channel state information (CSI) at the transmitter and receiver, which is usually difficult to obtain in mmWave communications. To be specific, the high cost and power consumption of mixed-signal hardware with high sampling rates limit the number of radio-frequency (RF) chains in practical mmWave transceivers \cite{RH-Survey}, making fully-digital processing techniques such as the traditional least-squared method impractical \cite{EST-OMP}. On the other hand, mmWave channels usually have a limited number of propagation paths, and therefore it is sufficient to only estimate channel parameters of these paths, which include angle of arrivals (AoAs), angle of departures (AoDs), and propagation gains.

There have been extensive work on channel estimation or beam training for mmWave systems over the years \cite{Beam-CB, RH-MSHC, Xiao-TWC, DJ-TWC, Matt-TSP, Matt-TCOM, Garcia-Tracking}. A widely used channel estimation method is to sequentially transmit highly directional training beams steering to different directions over time and pick the direction with the largest received signal-to-noise ratio (SNR) \cite{Beam-CB}. Nevertheless, this method is time-consuming and the required number of training beams is usually large in order to achieve a favorable estimate of the channel. This problem becomes even more challenging in mobile scenarios, where the channel keeps changing and the transmitter needs to frequently send training beams to update the estimation results, increasing the training overhead considerably. Therefore, an efficient and accurate beam training strategy is crucially important in mobile scenarios. In \cite{RH-MSHC, Xiao-TWC, DJ-TWC}, adaptive compressed sensing (CS) algorithms have been used to estimate the mmWave channels, which essentially search potential paths using hierarchical multi-resolution codebooks. These adaptive CS algorithms can achieve a favorable estimation performance but require excessive feedback, which may significantly exacerbate the system overhead. Meanwhile, all the aforementioned channel estimation algorithms fail to capture the temporal correlation between consecutive channel realizations in mobile mmWave communications.

Recently, it has been shown in \cite{He-Tracking, Zhang-Tracking, RH-Tracking, Gao-TWC, Palacios-Tracking, Li-ICASSP} that the temporal correlation between channel realizations can be exploited to further improve the beam training efficiency. Specifically, the CSI in current channel realization is closely related to that of the previous one, and this relationship can be used to speed up the beam training procedure. This type of priori-aided beam training technique is referred to as channel tracking or beam tracking in the literature \cite{He-Tracking, Zhang-Tracking, RH-Tracking, Gao-TWC}. To date, most existing works that investigate beam tracking techniques for mmWave channels have focused on the assumption that the values of AoA and AoD vary smoothly. In \cite{Zhang-Tracking, RH-Tracking, Gao-TWC, Palacios-Tracking, Li-ICASSP}, the temporal variation of AoA/AoD over the considered period of time is assumed to follow a Markov process, and the AoA's and AoD's deviations between two consecutive channel realizations are modeled as small Gaussian random variables, based on which various Kalman filter-based beam tracking algorithms have been developed. It is also worth mentioning that the authors in \cite{Location-Aided, Dai-Tracking, Sina-Tracking, Zang-Handover} have proposed to employ the mobile users' location and trajectory information to reduce the beam training overhead. However, these strategies are limited to vehicular networks and not universal.

To incorporate the abrupt changes of mmWave channels due to blockage or other environmental obstructions, several works have proposed to employ the discrete Markov process to model the temporal variations of AoA/AoD \cite{Duan-Tracking, My-ICCW, ML, MAP, POMDP}. It has been shown in \cite{Duan-Tracking} that the problem of tracking the time-varying AoA and AoD can be transformed into finding the support of the sparse beamspace channel, which is solved by invoking an approximate message passing algorithm. In \cite{My-ICCW, ML, MAP, POMDP}, codebook-based training beamforming vectors (beams) are adopted to reduce the design complexity. Specifically, a set of codewords (a beam codebook consists of a sequence of codewords and each codeword is a beamforming vector steering to a specific direction) that can minimize the Cramer-Rao lower bound averaged over the priori distribution of AoD are selected for beam tracking in \cite{ML} and \cite{MAP}, and the maximum likelihood (ML) and maximum a posteriori (MAP) criteria are respectively used to estimate the true direction of AoD. In \cite{POMDP}, the beam tracking problem is equivalent to a partially observable Markov decision process (POMDP), where the selected training beams serving as actions of the POMDP depend on the belief vector, observation information and reward. In the high SNR regime, the observations of the beam (angular) space based on signal detection are reliable, and hence the selections of training beams using this POMDP framework are appropriate. However, the consequent high power consumption will significantly increase the dynamic range of the power amplifiers, considerably increasing the hardware cost. On the other hand, when SNR goes low, the observations of the beam space will be inaccurate, and therefore selecting the optimal training beam sequence based on this POMDP framework are unreliable and the true AoAs/AoDs can be lost.

To address the aforementioned problems, we develop a new beam pair allocation strategy aiming to maximize the average successful tracking probability (ASTP) of time-varying mmWave multiple-input multiple-output (MIMO) systems, which can work effectively for all SNR regimes. Motivated by \cite{ML, MAP, POMDP}, the temporal variations of AoA and AoD within the considered period of time are modeled as two discrete Markov processes, described by their associated transition probabilities respectively, which are assumed to be known. Highly directional transmitting (Tx) and receiving (Rx) training beams picked from two predefined codebook matrices are used to combat the severe propagation loss. To further increase the received signal powers in specific directions and consequently improve the beam tracking performance, we allow the Tx-Rx beam pairs steering to these directions to be used repeatedly in the beam training period, which is different from most existing works\footnote{It is worth mentioning that the Tx-Rx training beam pairs are also allowed to be used repeatedly in \cite{MAB}. However, since an online stochastic optimization known as multi-armed bandit algorithm is considered, one feedback of the received energy is needed after each measurement, which may significantly exacerbate the system overhead.}. In the following, we summarize the methodologies and main contributions of this paper:

1) In the case of orthogonal Tx-Rx beam pairs, a power-based estimator that returns the direction with the largest received signal power is employed to track the time-varying AoA and AoD, leading to a closed-form expression for the (one-step) ASTP. Since the number of repetitions of each Tx-Rx beam pair can only be an integer, selecting the optimal set of the Tx-Rx beam pairs and determining their associated repetition times to maximize the ASTP is shown to be equivalent to an integer nonlinear programming (I-NLP). Though determining the optimal Tx-Rx beam pair sequence is NP-hard, we prove that the Tx-Rx beam pair with a higher transition probability should be used more times than those with lower transition probabilities in order to achieve the maximal ASTP.

2) It is very challenging to optimize the exact repetition times of each Tx-Rx beam pair due to the complicated expression for the ASTP. Therefore, we derive a tractable approximation for the ASTP as the new objective function of the formulated I-NLP. Afterward, we divide its feasible region into a set of subregions and construct a series of concave sub I-NLPs, which can be solved via recursively invoking a nonlinear branch-and-bound (N-BB) algorithm. To avoid the computational cost of the iterative N-BB algorithm, we relax the integer constraints of each sub I-NLP and obtain a promising solution by solving the Karush-Kuhn-Tucker (KKT) conditions of these relaxed subproblems following a similar recursive manner as in the iterative N-BB algorithm.

3) For the more general scenario where the Tx-Rx training beam pairs are overlapped in the angular space, the power-based estimator performs poorly due to the non-negligible inter-beam interference. In this case, we modify the power-based estimator and propose to track the time-varying AoA and AoD via an orthogonal matching pursuit (OMP) algorithm, which essentially exploits the inter-beam interference to improve the ASTP. Moreover, since no explicit expression for the ASTP exists when the OMP-based estimator is adopted, a closed-form lower bound of the ASTP can be derived, based on which a favorable beam pair allocation strategy to maximize the ASTP is obtained. Our numerical results demonstrate the superiority of the proposed beam pair allocation strategy over the uniform and proportional allocation strategies.

The rest of the paper is organized as follows. In Section \ref{SM}, we describe the considered mmWave system model and the adopted beam training protocol. In Section \ref{S-OBP}, orthogonal Tx-Rx training beam pairs are assumed and a power-based estimator is used to track the time-varying AoA and AoD. In Section \ref{S-NOBP}, we consider a more general scenario where the Tx-Rx training beam pairs are overlapped in the angular space. Numerical results are provided in Section \ref{NR}, followed by the conclusions in Section \ref{CN}.

\textbf{Notations}: Bold uppercase $\bf A$ and lowercase $\bf a$ represent matrices and column vectors respectively, and non-bold letters are scalars. ${\bf A}^{\ast}$, ${\bf A}^{\rm T}$, and ${\bf A}^{\rm H}$ represent the conjugate, transpose, conjugate transpose of $\bf A$, respectively. ${\bf A}[m, :]$, ${\bf A}[:, n]$ and ${\bf A}[m, n]$ are the $m$-th row, the $n$-th column, and the $(m, n)$-th element of $\bf A$, respectively. ${\bf A} \otimes {\bf B}$ is the Kronecker product of $\bf A$ and $\bf B$. $(b_1 \bullet b_2)_{N} \triangleq b_1 + N (b_2 - 1)$. ${\cal {CN}}({\bf a}, {\bf A})$ denotes a complex Gaussian distribution with mean $\bf a$ and covariance matrix $\bf A$. ${\bf I}$ is the identity matrix. ${\bf a} \triangleq \text{vec}({\bf A})$ is the vectorization operation by stacking the columns of $\bf A$ into a vector $\bf a$. $\big\{a_1, \cdots, a_n\big\} \big \backslash \big\{b_1, \cdots, b_k\big\}$ represents the set $\big\{a_1, \cdots, a_n\big\}$ excluding $\big\{b_1, \cdots, b_k\big\}$. $\displaystyle \binom{n}{k}$ is the number of $k$-combinations of an $n$-element set. $\exp$ is the exponential function. $\Pr\big\{S_2 = b_2 \leftarrow S_1 = b_1\big\}$ represents the transition probability from $S_1 = b_1$ to $S_2 = b_2$, while $\Pr\big(S_2 = b_2 \mid S_1 = b_1 \big)$ is the probability of $S_2 = b_2$ conditioned on $S_1 = b_1$. ${\mathds N}^{+}$ and ${\mathds N}^{++}$ denote the nonnegative integer set and positive integer set, respectively.

\section{System Model} \label{SM}
We consider a mmWave MIMO system, in which a base station (BS) equipped with $N_{\rm T}$ antennas communicates with a mobile station (MS) equipped with $N_{\rm R}$ antennas. Denote the $N_{\rm T} \times 1$ transmitting beamforming vector and the $N_{\rm R} \times 1$ receiving beamforming vector by $\bf f$ and $\bf w$ respectively, which are normalized to satisfy $\|{\bf f}\|^2 = \|{\bf w}\|^2 = 1$. Moreover, the pilot symbol is denoted by $x = \sqrt{P}$, where $P$ is the power consumed per transmission in the beam training period, and the received signal is then written as
\begin{equation}\label{SM-1}
    {\bf r} = {\bf H}{\bf f} x + \tilde {\bf n},
\end{equation}
where $\bf H$ is the $N_{\rm R} \times N_{\rm T}$ channel matrix between the MS and BS, and $\tilde {\bf n}$ is the $N_{\rm R} \times 1$ complex additive white Gaussian noise, i.e., $\tilde {\bf n} \sim {\cal CN}({\bm 0}, \sigma_0^2 {\bf I})$. The MS adopts the receiving beamforming vector $\bf w$ to process the received signal $\bf r$, given by
\begin{equation}\label{SM-2}
    y = {\bf w}^{\rm H} {\bf r} = {\bf w}^{\rm H} {\bf H}{\bf f} x + n,
\end{equation}
where $n = {\bf w}^{\rm H} \tilde {\bf n}$ and $n \sim {\cal CN}(0, \sigma_0^2)$ due to $\|{\bf w}\|^2 = 1$.

\subsection{Millimeter-Wave Channel} \label{MWCM}
Since mmWave channels have a very limited number of scatters, as in \cite{Sayeed}, we use the geometric channel model to express $\bf H$ as
\begin{equation} \label{SM-3}
    {\bf H} = \sqrt{N_{\rm T} N_{\rm R}} \sum\limits_{\ell = 1}^{L} \alpha_\ell {\bf a}_{\rm R}(\theta_\ell) {\bf a}_{\rm T}^{\rm H} (\vartheta_\ell),
\end{equation}
where $L$ is the total number of propagation paths, $\alpha_\ell, \theta_\ell, \vartheta_\ell$ are the complex gain, the normalized AoA and AoD of the $\ell$-th path respectively, and $\alpha_\ell \sim {\cal CN}(0, \sigma_\alpha^2)$, $\forall \ell = 1, \cdots, L$. In addition, ${\bf a}_{\rm R}(\theta_\ell)$ and ${\bf a}_{\rm T}(\vartheta_\ell)$ are termed antenna array response vectors. In this paper, we assume that uniform linear arrays (ULAs) are used at the MS and BS, and therefore ${\bf a}_{\rm R}(\theta_\ell)$ and ${\bf a}_{\rm T}(\vartheta_\ell)$ can be respectively written as
\begin{eqnarray*}
    {\bf a}_{\rm R}(\theta_\ell) & = & \frac{1}{\sqrt{N_{\rm R}}} \big[1, ~e^{j \theta_\ell}, ~\cdots, ~e^{j (N_{\rm R} - 1) \theta_\ell} \big]^{\rm T}, \label{SM-4A} \\
    {\bf a}_{\rm T}(\vartheta_\ell) & = & \frac{1}{\sqrt{N_{\rm T}}} \big[1, ~e^{j \vartheta_\ell}, ~\cdots, ~e^{j (N_{\rm T} - 1) \vartheta_\ell} \big]^{\rm T}, \label{SM-4B}
\end{eqnarray*}
where $j$ is the imaginary unit, i.e., $j = \sqrt{-1}$. Moreover, the relationship between the normalized AoA (AoD) and the physical AoA (AoD) is expressed as \cite{POMDP}
\begin{equation*}
    \theta_\ell = \frac{2 \pi d \sin(\tilde \theta_\ell)} {\lambda_s}, ~~\vartheta_\ell = \frac{2 \pi d \sin(\tilde \vartheta_\ell)} {\lambda_s}, ~\forall \ell = 1, \cdots, L,
\end{equation*}
where $\lambda_s$ is the signal wavelength, $d$ is the distance between two adjacent antenna elements, $\tilde \theta_\ell$ and $\tilde \vartheta_\ell$ are the physical AoA and AoD of the $\ell$-th path respectively. By letting $d = \frac{\lambda_s}{2}$, we obtain that $\theta_\ell, \vartheta_\ell \in [-\pi, \pi]$ when $\tilde \theta_\ell, \tilde \vartheta_\ell \in [-\pi, \pi]$.

\subsection{Time-Varying AoA and AoD} \label{DCM}
In order to exploit the sparsity of the mmWave \mbox{channels, as} in \cite{RH-MSHC, Matt-TSP, MAP, POMDP}, we introduce two beam codebook matrices ${\bf A}_{\rm R} = \big[{\bf a}_{\rm R}(\bar \theta_1), {\bf a}_{\rm R}(\bar \theta_2), \cdots, {\bf a}_{\rm R}(\bar \theta_{X_{\rm R}})\big]$ and ${\bf A}_{\rm T} = \big[{\bf a}_{\rm T}(\bar \vartheta_1), {\bf a}_{\rm T}(\bar \vartheta_2), \cdots, {\bf a}_{\rm T}(\bar \vartheta_{X_{\rm T}})\big]$, where
\begin{eqnarray*}
    \bar \theta_m & = & \frac{2 \pi (m - 1)}{X_{\rm R}} - \frac{\pi(X_{\rm R} - 1)}{X_{\rm R}}, ~m = 1, \cdots, X_{\rm R}, \\
    \bar \vartheta_n & = & \frac{2 \pi (n - 1)}{X_{\rm T}} ~- \frac{\pi(X_{\rm T} - 1)}{X_{\rm T}}, ~n~ = 1, \cdots, X_{\rm T},
\end{eqnarray*}
to divide the whole angular space $[-\pi, \pi]$ into $X_{\rm R}$ and $X_{\rm T}$ directions, respectively. In addition, we follow \cite{RH-Survey, RH-MSHC, POMDP} and assume that $\{\theta_\ell\}$ and $\{\vartheta_\ell\}$ are respectively taken from the sets $\{\bar \theta_m\}_{m = 1}^{X_{\rm R}}$ and $\{\bar \vartheta_n\}_{n = 1}^{X_{\rm T}}$ for simplicity\footnote{Though this on-grid assumption of the normalized AoAs/AoDs may not be rigorous, the resulting quantization errors are not significant when $X_{\rm R}$ and $X_{\rm T}$ are large enough \cite{RH-Survey, RH-MSHC, POMDP}. The off-grid case where the values of $\{\theta_\ell\}$ and $\{\vartheta_\ell\}$ are continuous is left as one of our future works.}.

\begin{figure}[!t]
    \centering
    \includegraphics[width = 16.8cm]{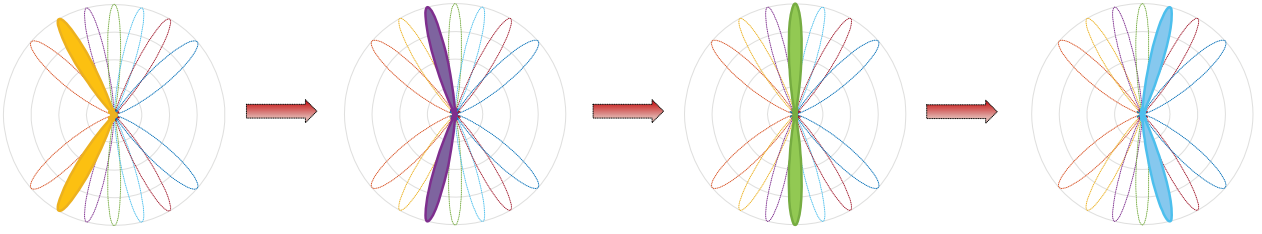}
    \caption{An example of the temporal variations of one AoA (AoD).} \label{AoA}
\end{figure}

In mobile scenarios, as mentioned earlier, the channel realizations between two consecutive transmission blocks are correlated. Following \cite{ML, MAP, POMDP}, we model the temporal variations of each AoA (AoD) within a set of transmission blocks as a discrete Markov process (see Fig. \ref{AoA} and Fig. \ref{FRAME}), described by the following transition probability
\begin{equation} \label{SM-5A}
    \Pr\left\{\theta_\ell^{[\tau]} = \bar \theta_{k_1} \leftarrow \theta_\ell^{[\tau - 1]} = \bar \theta_{k_0} \right\} = C \beta^{|k_1 - k_0|},
\end{equation}
where $k_0, k_1 \in \big\{1, \cdots, X_{\rm R}\big\}$. The superscript $\tau$ denotes the $\tau$-th beam training period or transmission block, where $\tau = 2, \cdots, T$. The variable $\beta \in [0, 1]$ indicates the variation speed of the AoAs, and $C$ is the normalization coefficient. According to \eqref{SM-5A}, when $\beta$ is small, e.g., $\beta = 0.1$, the updated AoA $\theta_\ell^{[\tau]}$ is very likely to be in the proximity to $\theta_\ell^{[\tau-1]}$. On the other hand, when $\beta = 1.0$, the AoAs change rapidly and $\theta_\ell^{[\tau]}$ will be uniformly distributed in the set $\{\bar \theta_m\}_{m = 1}^{X_{\rm R}}$, which corresponds to the abrupt changes in mmWave channels. The associated transition probability of each AoD can be similarly expressed as in \eqref{SM-5A}, given by
\begin{equation} \label{SM-5B}
    \Pr\left\{\vartheta_\ell^{[\tau]} = \bar \vartheta_{i_1} \leftarrow \vartheta_\ell^{[\tau - 1]} = \bar \vartheta_{i_0} \right\} = \tilde C \tilde \beta^{|i_1 - i_0|},
\end{equation}
where $i_0, i_1 \in \big\{1, \cdots, X_{\rm T}\big\}$ and $\tilde \beta$ is introduced to indicate the variation speed of the AoDs. Moreover, the channel gains $\{\alpha_\ell\}$ are assumed to change independently from one transmission block to another \cite{POMDP}. Finally, it is worth highlighting that the proposed beam pair allocation strategy can be extended to other types of transition probabilities such as the ones used in \cite{ML} and \cite{MAP}.

\subsection{Beam Training Protocol}
\begin{figure}[!t]
    \centering
    \includegraphics[width = 12.8cm]{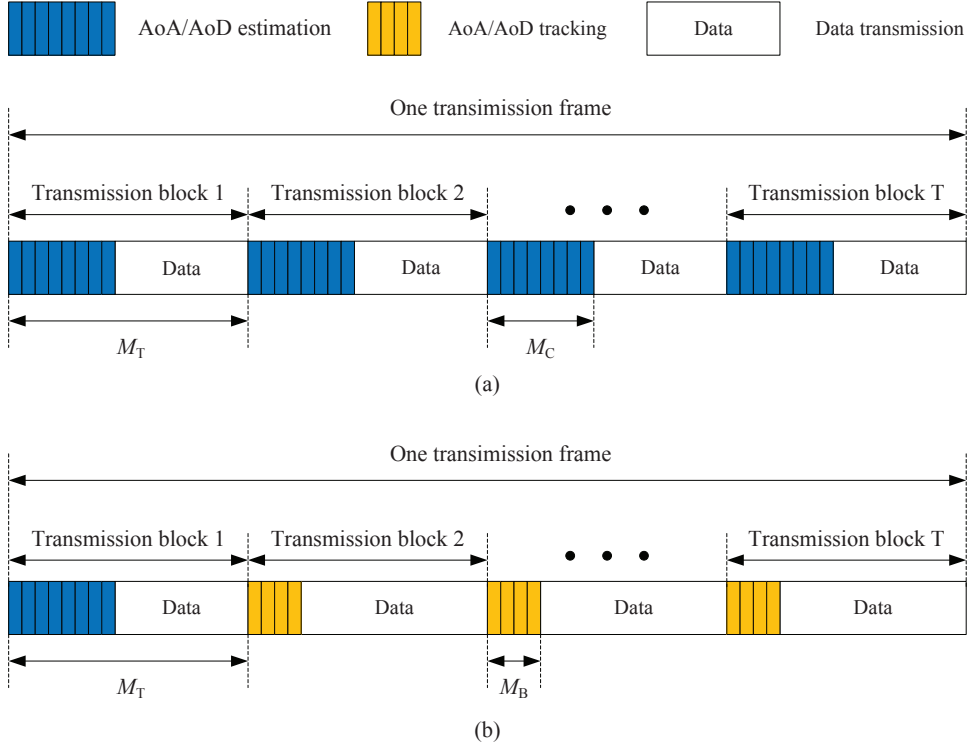}
    \caption{Frame structure of (a) conventional beam training and (b) adopted beam training.} \label{FRAME}
\end{figure}

To track the time-varying AoAs and AoDs, which can result from the mobility of the MS or the reflection scatters, the BS transmits a sequence of pilot symbols via using a set of dedicated training beams to the MS periodically, which also uses a set of dedicated beams to receive them in different directions. As shown in Fig. \ref{FRAME}, one transmission frame is assumed to consist of $T$ transmission blocks, and each transmission block is made up of $M_{\rm T}$ symbol durations, where the first $M_{\rm C}$ or $M_{\rm B}$ symbol durations are used for beam training and the rest are for data communication. For the conventional beam training protocol shown in Fig. \ref{FRAME}(a), as no priori information is used, in each beam training period the BS and MS consume a fixed number of $M_{\rm C}$ beams in $M_{\rm C}$ symbol durations to estimate the AoAs and AoDs. While for the adopted beam training protocol depicted in Fig. \ref{FRAME}(b), traditional channel estimation is performed in the first transmission block, since no priori information can be exploited at this block. In each of the subsequent transmission blocks, $M_{\rm B}$ training beams in $M_{\rm B}$ symbol durations are selected to execute beam tracking based on the previous estimate\footnote{When one beam training period has finished, the MS feeds back the estimated AoAs and AoDs to the BS for the subsequent data transmission and the next beam tracking procedure.} and the priori transition probabilities, which will be introduced later. Moreover, as $M_{\rm B} < M_{\rm C}$, a larger fraction of time can be left for data communication, leading to a higher throughput. Finally, since channel estimation techniques have been widely investigated in the existing literature \cite{Beam-CB, RH-MSHC, Xiao-TWC, DJ-TWC, Matt-TSP}, we only consider the beam tracking strategy commencing from the second beam training period.

In the rest of the paper, we focus on the $\tau$-th beam training period unless otherwise specified, $\exists \tau \in \{2, \cdots, T\}$, such that $\theta_\ell^{[\tau - 1]} = \bar \theta_{k_0}$ and $\theta_\ell^{[\tau]} = \bar \theta_{k_1}$ are the previous and the current AoAs of the $\ell$-th path. In accordance with \eqref{SM-5B}, the previous AoD and the current AoD of the $\ell$-th path are represented by $\vartheta_\ell^{[\tau - 1]} = \bar \vartheta_{i_0}$ and $\vartheta_\ell^{[\tau]} = \bar \vartheta_{i_1}$, respectively. Moreover, while below we only consider a single-path channel model and drop the subscript $\ell$ for the sake of convenience, the proposed beam tracking strategy can be readily extended to the multi-path scenario, which is discussed at the end of Section \ref{S-NOBP}.

\section{Special Case: Unitary Codebook Matrices} \label{S-OBP}
In this section, to gain some insights, we follow \cite{POMDP} and assume that $X_{\rm T} = N_{\rm T}$ and $X_{\rm R} = N_{\rm R}$, such that the two beam codebook matrices ${\bm A}_{\rm T}$ and ${\bm A}_{\rm R}$ become two discrete Fourier transformation (DFT) matrices.

\subsection{Closed-Form ASTP} \label{ASTP}
Similar to \cite{ML, MAP, POMDP}, we pick the Tx and Rx training beams from ${\bf A}_{\rm T}$ and ${\bf A}_{\rm R}$, respectively, and when ${\bf f}^{[m]} = {\bf a}_{\rm T}(\bar \vartheta_i)$ and ${\bf w}^{[m]} = {\bf a}_{\rm R}(\bar \theta_k)$ are chosen at the $m$-th symbol duration or measurement, $\exists m = 1, \cdots, M_{\rm B}$, the received symbol can be given by
\begin{eqnarray}\label{ATEP-1}
    y_{k, i} & = & ({\bf w}^{[m]})^{\rm H} {\bf H} {\bf f}^{[m]} x^{[m]} ~~~~+~~~ ({\bf w}^{[m]})^{\rm H} \tilde {\bf n}^{[m]} \nonumber \\[5pt]
    & = & \gamma \alpha {\bf a}^{\rm H}_{\rm R}(\bar \theta_k) {\bf a}_{\rm R}(\bar \theta_{k_1}) {\bf a}^{\rm H}_{\rm T}(\bar \vartheta_{i_1}) {\bf a}_{\rm T}(\bar \vartheta_i) + n^{[m]} \nonumber \\ [5pt]
    & \overset{(a)}{=} & \begin{dcases}
    \gamma \alpha + n^{[m]}, & \text{if}~ k = k_1 ~\text{and}~ i = i_1, \\
    n^{[m]}, & \text{otherwise},
    \end{dcases}
\end{eqnarray}
where $\gamma = \sqrt{P N_{\rm T} N_{\rm R}}$ and (a) is due to the fact that ${\bf A}_{\rm T}$ and ${\bf A}_{\rm R}$ are two unitary matrices.

In order to improve the received signal power in a specific direction, its associated Tx-Rx beam pair is allowed to be used repeatedly. For convenience, we use ${\cal B}_{k, i}$ to denote the Tx-Rx beam pair ${\bf w} = {\bf a}_{\rm R}(\bar \theta_k)$ and ${\bf f} = {\bf a}_{\rm T} (\bar \vartheta_i)$, and ${\cal B} \triangleq \big\{{\cal B}_{k, i} \mid 1 \le k \le X_{\rm R}, 1 \le i \le X_{\rm T}\big\}$ is the set consisting of all potential beam pairs. Moreover, the repetition times of ${\cal B}_{k, i}$ during one beam training period is denoted by $\lambda_{k, i}$, and the corresponding received symbols are expressed as ${\bf y}_{k, i}[1], \cdots, {\bf y}_{k, i}[\lambda_{k, i}]$, respectively. By adding up the $\lambda_{k, i}$ received symbols, we obtain that
\begin{eqnarray}\label{ATEP-2}
    \xi_{k, i} = \sum\limits_{m = 1}^{\lambda_{k, i}} {\bf y}_{k, i}[m] \sim  \begin{dcases}
    {\cal CN}\big(\gamma \alpha \lambda_{k, i}, \sigma_0^2 \lambda_{k, i} \big), & \text{if}~ k = k_1 ~\text{and}~ i = i_1, \\
    {\cal CN}\big(0, \sigma_0^2 \lambda_{k, i}\big), & \text{otherwise}. \\
    \end{dcases}
\end{eqnarray}

Recall that there are in total $X \triangleq X_{\rm T} \times X_{\rm R}$ distinct beam pairs in ${\cal B}$. For notational simplicity, we redefine ${\cal B}_n$ to represent the $n$-th beam pair, $\forall n = 1, \cdots, X$, and obviously a one-to-one mapping exists between ${\cal B}_n$ and ${\cal B}_{k, i}$, which is denoted by $n = (k \bullet i)_{X_{\rm R}}$. Accordingly, in the rest of the paper, we use $\xi_n$ and $\lambda_n$ to replace $\xi_{k, i}$ and $\lambda_{k, i}$, respectively.

In the current beam training period, without loss of generality, we denote the selected Tx-Rx training beam pairs by ${\cal B}_{z_1}, \cdots, {\cal B}_{z_N}$, where $z_1, \cdots, z_N \in \{1, \cdots, X\}$ remain to be optimized with $N \le M_{\rm B}$ since one beam pair might be used repeatedly. Following \cite{Sayeed-PBE}, a power-based estimator is introduced to estimate the updated AoA and AoD for its simplicity, and when $z_n \triangleq (a_n \bullet c_n)_{X_{\rm R}} = (k_1 \bullet i_1)_{X_{\rm R}}$, the successful estimation probability is given by
\begin{eqnarray}\label{ATEP-3}
    \Gamma_{z_n, |\alpha|^2} = \Pr\Bigg(\bigcap\limits_{m = 1, m \ne n}^N |\xi_{z_n}|^2 > |\xi_{z_m}|^2 ~\Big |~ \alpha \Bigg).
\end{eqnarray}
It can be observed from \eqref{ATEP-2} that $|\xi_{z_n}|^2$ satisfies a non-central chi-squared distribution while $|\xi_{z_m}|^2$ follows an exponential distribution, $\forall m \ne n$, and therefore we can rewrite \eqref{ATEP-3} as
\begin{eqnarray}\label{ATEP-4}
    \Gamma_{z_n, |\alpha|^2} = \int\limits_0^\infty h \bigg(u; \lambda_{z_n}, |\alpha|^2 \bigg) \prod\limits_{m = 1, m \ne n}^N \left(1 - \exp \left(-\frac{u}{\lambda_{z_m} \sigma_0^2}\right) \right) \text{d} u,
\end{eqnarray}
where $h(u; \lambda_{z_n}, |\alpha|^2)$ is given by
\begin{eqnarray}\label{ATEP-5}
    h\Big(u; \lambda_{z_n}, |\alpha|^2\Big) = \frac{1} {\lambda_{z_n} \sigma_0^2} \exp\left(-\frac{u + \lambda^2_{z_n} \gamma^2 |\alpha|^2} {\lambda_{z_n} \sigma_0^2}\right) I_0\left(\frac{\sqrt{4 \gamma^2 |\alpha|^2 u}} {\sigma_0^2}\right),
\end{eqnarray}
and $I_0(\cdot)$ is the zero-th order modified Bessel function of the first kind. Moreover, we need to integrate $\Gamma_{z_n, |\alpha|^2}$ over the exponential distribution of $|\alpha|^2$, which is expressed as
\begin{eqnarray}\label{ATEP-7}
    \Gamma_{z_n} & = & \int\limits_0^\infty \int\limits_0^\infty ~~h \bigg(u; \lambda_{z_n}, |\alpha|^2 \bigg)~\prod\limits_{m = 1, m \ne n}^N \left(1 - \exp\left(-\frac{u}{\lambda_{z_m} \sigma_0^2} \right) \right) ~\frac{1}{\sigma_\alpha^2} ~\exp \left(-\frac{|\alpha|^2} {\sigma_\alpha^2}\right) ~\text{d}u ~\text{d}|\alpha|^2 \nonumber \\
    & = & \int\limits_0^\infty~\frac{1}{\lambda^2_{z_n} \gamma^2 \sigma_\alpha^2 + \lambda_{z_n} \sigma_0^2} \exp\left(-\frac{u} {\lambda^2_{z_n} \gamma^2 \sigma_\alpha^2 + \lambda_{z_n} \sigma_0^2}\right) \prod\limits_{m = 1, m \ne n}^N \left(1 - \exp\left(-\frac{u}{\lambda_{z_m} \sigma_0^2} \right) \right) \text{d} u \nonumber \\
    & = & 1 - \sum\limits_{\kappa_1 = 1}^{N - 1} (-1)^{\kappa_1 + 1} \sum\limits_{\kappa_2 = 1}^{\binom{N - 1}{\kappa_1}} \frac{1}{1 + \sum\nolimits_{\kappa_3 = 1}^{\kappa_1} \big(\lambda^2_{z_n} \gamma^2 \sigma_\alpha^2 + \lambda_{z_n} \sigma_0^2\big) ~\big /~ \big(\sigma_0^2 \lambda_{\kappa_1, \kappa_2, \kappa_3, -z_n} \big)},
\end{eqnarray}
where $\lambda_{\kappa_1, \kappa_2, \kappa_3, -z_n} \in \{\lambda_{z_1}, \cdots, \lambda_{z_N}\} \big \backslash \{\lambda_{z_n}\}$.

The (one-step) ASTP can be therefore expressed as
\begin{equation}\label{ATEP-8}
    \bar \Gamma_1 = \sum\limits_{n = 1}^N \pi_{z_n} \times \Gamma_{z_n},
\end{equation}
where $\pi_{z_n}$ denotes the transition probability from $\Big\{\bar \theta_{k_0}, \bar \vartheta_{i_0}\Big\}$ to $\Big\{\bar \theta_{a_n}, \bar \vartheta_{c_n}\Big\}$, which is directly calculated from \eqref{SM-5A} and \eqref{SM-5B}.

\subsection{Problem Formulation} \label{I-NLP}
In this paper, we aim to seek the optimal Tx-Rx training beam pairs ${\cal B}_{z_1}, \cdots, {\cal B}_{z_N}$ and their associated repetition times ${\bm \lambda} = [\lambda_{z_1}, \cdots, \lambda_{z_N}]^{\rm T}$ that can maximize the ASTP, given the total number of pilot symbol durations $M_{\rm B}$. To this end, the following optimization problem is formulated:
\begin{eqnarray}
    \textbf{(P1)} & \max\limits_{\bm \lambda} & \bar \Gamma_1(\bm \lambda) \label{P1-A} \\
    & {\rm s.t.} & \lambda_{z_1} + \cdots + \lambda_{z_N} ~= M_{\rm B}, \label{P1-B} \\
    && \lambda_{z_1}, \lambda_{z_2}, \cdots, \lambda_{z_N} \in \mathds N^{++}, \label{P1-C} \\
    && z_1, ~z_2, ~\cdots, ~z_N \in \{1, \cdots, X\}. \label{P1-D}
\end{eqnarray}
It is observed that \textbf{(P1)} is an integer nonlinear programming problem, which is in general NP-hard. In the subsequent theorem, we show that the domain of \textbf{(P1)} can be substantially reduced.
\begin{theorem}\label{theorem1}
    If the $X$ possible beam pairs in $\cal B$ are sorted in a descending order according to their associated transition probabilities, $\{{\cal B}_1, \cdots, {\cal B}_X\} \rightarrow \{{\cal B}_{s_1}, \cdots, {\cal B}_{s_X}\}$, in order to achieve the optimal ASTP, the numbers of used Tx-Rx beam pairs should satisfy $\lambda_{s_1} \ge \cdots \ge \lambda_{s_X}$.
\end{theorem}
\begin{IEEEproof}
    Refer to Appendix \ref{theorem-proof}.
\end{IEEEproof}

Thanks to \textbf{Theorem 1}, in the following we can use $\{{\cal B}_{s_1}, \cdots, {\cal B}_{s_N}\}$ to replace $\{{\cal B}_{z_1}, \cdots, {\cal B}_{z_N}\}$ and rewrite \textbf{(P1)} as
\begin{eqnarray}
    \hspace{-0.92cm} \textbf{(P2)} & \max\limits_{\bm \lambda} & \bar \Gamma_1 (\bm \lambda) = \sum\limits_{n = 1}^N \pi_{s_n} \times \Gamma_{s_n} (\bm \lambda) \label{P2-A} \\
    \hspace{-0.92cm} & {\rm s.t.} & \lambda_{s_1} + ~\cdots~ + \lambda_{s_N} ~= M_{\rm B}, \label{P2-B} \\
    \hspace{-0.92cm} & & \lambda_{s_1}, ~\lambda_{s_2}, \cdots, ~\lambda_{s_N} \in ~\mathds{N}^{++}. \label{P2-C}
\end{eqnarray}
It is still challenging to handle \textbf{(P2)} due to the complicated structure of its objective function in \eqref{P2-A}, and therefore we first simplify it into a more tractable form.
\begin{lemma}\label{lemma2}
    If ${\cal B}_{s_1}, \cdots, {\cal B}_{s_N}$ are used in the beam training period, the ASTP with the power-based estimator is lower bounded by
    \begin{equation}
        \bar \Gamma_1^{\rm lb} (\bm \lambda) = \sum\limits_{n = 1}^N \pi_{s_n} \left[1 - \frac{M_{\rm B} - \lambda_{s_n}} { \lambda_{s_n}^2 r_0 + \lambda_{s_n}}\right],
    \end{equation}
    and upper bounded by
    \begin{equation}
        \bar \Gamma_1^{\rm ub} (\bm \lambda) = \sum\limits_{n = 1}^N \pi_{s_n} \left[1 - \frac{{\cal F}(N - 1)}{\lambda_{s_n}^2 r_0 + \lambda_{s_n}}\right].
    \end{equation}
    Furthermore, $\bar \Gamma_1(\bm \lambda)$ can be approximated by
    \begin{equation}
        \bar \Gamma_1^{\rm apx} (\bm \lambda) = \sum\limits_{n = 1}^N \pi_{s_n} \left[1 - \frac{{\cal F}(N - 1)}{N - 1} \frac{M_{\rm B} - \lambda_{s_n}}{\lambda_{s_n}^2 r_0 + \lambda_{s_n}} \right],
    \end{equation}
    or equivalently
    \begin{equation}
        \bar \Gamma_1^{\rm apx} (\bm \lambda) = \sum\limits_{n = 1}^N \pi_{s_n} \left[1 - \frac{M_{\rm B} - \lambda_{s_n}}{N - 1} \frac{{\cal F}(N - 1)}{\lambda_{s_n}^2 r_0 + \lambda_{s_n}} \right],
    \end{equation}
    where $\displaystyle r_0 = \frac{P N_{\rm T} N_{\rm R} \sigma_\alpha^2} {\sigma_0^2}$ and $\displaystyle {\cal F}(N - 1) = \sum\limits_{n = 1}^{N - 1}\frac{1}{n}$.
\end{lemma}
\begin{IEEEproof}
    Refer to Appendix \ref{lemma2-proof}.
\end{IEEEproof}

\subsection{Iterative Nonlinear Branch-and-Bound Algorithm}
Since ${\cal F}(N - 1) < N - 1$ and $M_{\rm B} - \lambda_{s_n} = \sum\nolimits_{m \ne n} \lambda_{s_m} > N - 1$, it is observed that
\begin{equation}\label{INLP-4}
    \bar \Gamma_1^{\rm lb} (\bm \lambda) < \bar \Gamma_1^{\rm apx} (\bm \lambda) < \bar \Gamma_1^{\rm ub} (\bm \lambda).
\end{equation}
In general, while we can pick any of $\bar \Gamma_1^{\rm lb}(\bm \lambda)$, $\displaystyle \bar \Gamma_1^{\rm ub}(\bm \lambda)$ and $\bar \Gamma_1^{\rm apx} (\bm \lambda)$ to replace $\bar \Gamma_1(\bm \lambda)$ as the new objective function, in the subsequent sections, we use $\bar \Gamma_1^{\rm apx}(\bm \lambda)$ and construct a new optimization problem, given by
\begin{eqnarray}\label{INLP-5}
    \textbf{(P3)} & \max\limits_{\bm \lambda} & \bar \Gamma_1^{\rm apx} (\bm \lambda) \\
    & {\rm s.t.} & \eqref{P2-B} ~\text{and}~ \eqref{P2-C}.
\end{eqnarray}
Though the exact value of $N$ in \eqref{INLP-5} is unknown, it has at most $M_{\rm B}$ cases, i.e., $N = 1, \cdots, M_{\rm B}$, and we can thus decompose \textbf{(P3)} into $M_{\rm B}$ subproblems, with each one corresponding to a specific $N$. By solving these subproblems, the optimal solution to \textbf{(P3)} can be obtained. Moreover, since these subproblems are concave I-NLPs as demonstrated in \textbf{Lemma \ref{lemma3}}, we can apply the nonlinear branch-and-bound method \cite{MINLP} to solve them optimally.

\begin{lemma}\label{lemma3}
    For a specific $N$, when we relax the integer variable $\lambda_{s_n}$ to a real variable $\tilde \lambda_{s_n}$, $\forall n = 1, \cdots, N$, $\bar \Gamma_1^{\rm apx}$ becomes a concave function with respect to $\tilde \lambda_{s_1}, \cdots, \tilde \lambda_{s_N}$.
\end{lemma}
\begin{IEEEproof}
    Refer to Appendix \ref{lemma3-proof}
\end{IEEEproof}

A closer observation of $\bar \Gamma_1^{\rm apx}$ shows that we may not need to solve all the $M_{\rm B}$ subproblems of \textbf{(P3)}. To be specific, when
\begin{eqnarray}\label{INLP-6}
    \bar \Gamma_1^{\rm apx} (\bm \lambda_N) = \sum\limits_{n = 1}^N \pi_{s_n} \left[1 - \frac{{\cal F}(N - 1)}{N - 1} \frac{M_{\rm B} - \lambda_{s_n}}{\lambda_{s_n}^2 r_0 + \lambda_{s_n}} \right] < \sum\limits_{n = 1}^N \pi_{s_n} \le \Delta,
\end{eqnarray}
where $\bm \lambda_N = [\lambda_{s_1}, \cdots, \lambda_{s_N}]^{\rm T}$ denotes a solution to the $N$-th subproblem, $\exists N \in \{2, \cdots, M_{\rm B}\}$, and $\Delta$ is a constant, we can see that
\begin{eqnarray}\label{INLP-7}
    \bar \Gamma_1^{\rm apx} (\bm \lambda_K) = \sum\limits_{n = 1}^K \pi_{s_n} \left[1 - \frac{{\cal F}(K - 1)}{K - 1} \frac{M_{\rm B} - \lambda_{s_n}}{\lambda_{s_n}^2 r_0 + \lambda_{s_n}} \right] < \sum\limits_{n = 1}^K \pi_{s_n} < \Delta,
\end{eqnarray}
where $K = 2, \cdots, N - 1$. In other words, when $\bar \Gamma_1^{\rm apx}(\bm \lambda_N) < \sum\nolimits_{n = 1}^N \pi_{s_n} \le \Delta$, the maximum value of $\bar \Gamma_1^{\rm apx}(\bm \lambda_K)$ will be less than $\Delta$ as well. By using this property, we can skip some subproblems to reduce the computational cost of \textbf{(P3)}. For clarity, the proposed iterative N-BB algorithm has been summarized in \textbf{Algorithm \ref{N-BB}}.

\begin{algorithm}[!htbp]
\caption{Proposed Iterative Nonlinear Branch-and-Bound Algorithm for \textbf{(P3)}} \label{N-BB}
\SetKwInOut{Input}{Input}\SetKwInOut{Output}{Output}
\Input{The total number of training beam pairs $M_{\rm B}$.}
\Output {The optimal solution to \textbf{(P3)}.}
Initialization: $\lambda_{s_n} = 1$, $\forall n = 1, \cdots, M_{\rm B}$, and $\Delta = \bar \Gamma_1^{\rm apx}(\bm \lambda_{M_{\rm B}})$. \\
Let $N = M_{\rm B} - 1$ and ${\cal T}(N) = \pi_{s_1} + \cdots + \pi_{s_N}$. \\
\While {${\cal T}(N) > \Delta$}
{
    Optimize $\bar \Gamma_1^{\rm apx}(\bm \lambda_N)$ via the N-BB method \cite{MINLP} and denote the maximal value by $\bar \Gamma_1^{\rm apx}(\bm \lambda_N^\star)$. \\
    Update the objective value: $\Delta = \bar \Gamma_1^{\rm apx}(\bm \lambda_N^\star)$ if $\bar \Gamma_1^{\rm apx}(\bm \lambda_N^\star) > \Delta$, otherwise, $\Delta$ remains unchanged. \\
    Let $N = N - 1$. \\
}
\end{algorithm}

Seen from \textbf{Algorithm 1}, when $N = M_{\rm B}$, the associated subproblem has only one solution, given by $\{\lambda_{s_n}\}_{n = 1}^{M_{\rm B}} = 1$, and its objective value $\bar \Gamma_1^{\rm apx}(\bm \lambda_{M_{\rm B}})$ is taken as a temporary lower bound of $\bar \Gamma_1^{\rm apx}(\bm \lambda)$. Next, we set $N = M_{\rm B} - 1$, which corresponds to $\lambda_{s_1} = 2$, $\{\lambda_{s_n}\}_{n = 1}^{M_{\rm B} - 1} = 1$, and $\lambda_{s_{M_{\rm B}}} = 0$. We evaluate $\bar \Gamma_1^{\rm apx} (\bm \lambda_{M_{\rm B} - 1})$ and compare it with the current lower bound $\bar \Gamma_1^{\rm apx} (\bm \lambda_{M_{\rm B}})$. If $\bar \Gamma_1^{\rm apx} (\bm \lambda_{M_{\rm B} - 1})$ is larger than $\bar \Gamma_1^{\rm apx} (\bm \lambda_{M_{\rm B}})$, we set $\bar \Gamma_1^{\rm apx} (\bm \lambda_{M_{\rm B} - 1})$ as the new lower bound of $\bar \Gamma_1^{\rm apx} (\bm \lambda)$. Otherwise, the current lower bound remains unchanged. We then consider $N = M_{\rm B} - 2$ by letting $\lambda_{s_{M_{\rm B} - 1}}, \lambda_{s_{M_{\rm B}}} = 0$, and $\{\lambda_{s_n}\}_{n = 1}^{M_{\rm B} - 2} \ge 1$. Optimize $\bar \Gamma_1^{\rm apx} (\bm \lambda_{M_{\rm B} - 2})$, compare its maximal value with the current lower bound and update the lower bound if applicable. This procedure is repeated until the temporary lower bound of $\bar \Gamma_1^{\rm apx} (\bm \lambda)$ cannot improve. The current lower bound is the global maximal value of $\bar \Gamma_1^{\rm apx} (\bm \lambda)$, and the corresponding solution is the optimal solution to \textbf{(P3)}.

\subsection{A Low-Complexity Solution} \label{S-LCS}
Though the iterative N-BB algorithm is able to solve \textbf{(P3)} optimally, its computational cost is high. In order to reduce the complexity, a suboptimal solution to \textbf{(P3)} is also provided by exploiting the KKT conditions. To be more specific, we first relax the $N$-th integer subproblem to a convex nonlinear optimization problem with respect to $\tilde \lambda_{s_1}, \cdots, \tilde \lambda_{s_N}$, given by
\begin{eqnarray}\label{LCS-1}
    \textbf{(P4)} & \min\limits_{\bm \lambda} & \sum\limits_{n = 1}^N \frac{\pi_{s_n}(M_{\rm B} - \tilde \lambda_{s_n})} {\tilde \lambda_{s_n}^2 r_0 + \tilde \lambda_{s_n}} \\
    & {\rm s.t.} & \tilde \lambda_{s_1} + \cdots + \tilde \lambda_{s_N} = M_{\rm B}, \\
    && \tilde \lambda_{s_1}, \tilde \lambda_{s_2}, \cdots, \tilde \lambda_{s_N} \ge 1.
\end{eqnarray}
The associated Lagrangian of \textbf{(P4)} is then expressed as
\begin{eqnarray}\label{LCS-2}
    {\cal L} = \sum\limits_{n = 1}^N \frac{\pi_{s_n} (M_{\rm B} - \tilde \lambda_{s_n})}{\tilde \lambda_{s_n}^2 r_0 + \tilde \lambda_{s_n}} + \mu_0 \left(\sum\limits_{n = 1}^N \tilde \lambda_{s_n} - M_{\rm B} \right) - \sum\limits_{n = 1}^N \mu_n \left(\tilde \lambda_{s_n} - 1 \right),
\end{eqnarray}
where $\mu_0, \mu_1, \cdots, \mu_N$ are the Lagrange multipliers, and the corresponding KKT conditions are given by
\begin{subequations}
    \begin{eqnarray} \label{LCS-3}
        \frac{\partial {\cal L}}{\partial \tilde \lambda_{s_n}} = \frac{\pi_{s_n}}{r_0} \left(\frac{1}{\tilde \lambda_{s_n}^2} - \frac{2 M_{\rm B}}{\tilde \lambda_{s_n}^3}\right) + \mu_0 - \mu_n = 0, ~\forall n = 1, \cdots, N, \\
        \mu_n (\tilde \lambda_{s_n} - 1) = 0, ~\mu_n \ge 0, ~\tilde \lambda_{s_n} \ge 1, ~\forall n = 1, \cdots, N, \\
        \sum\limits_{n = 1}^N \tilde \lambda_{s_n} = M_{\rm B}.
    \end{eqnarray}
\end{subequations}
It is worth mentioning that when we compute the partial derivative of $\cal L$ with respect to $\tilde \lambda_{s_n}$ in \eqref{LCS-3}, we approximate $\tilde \lambda_{s_n}^2 r_0 + \tilde \lambda_{s_n}$ by $\tilde \lambda_{\tau_n}^2 r_0$ since the SNR at the MS $r_0 = \frac{P N_{\rm T} N_{\rm R} \sigma_\alpha^2}{\sigma_0^2} \gg 1$. By solving the above KKT conditions, we can obtain that $\tilde \lambda_{s_n}^\star = \max\{1, b_n\}$, where
\begin{eqnarray}\label{LCS-4}
    b_n = \left(\frac{\pi_{s_n} M_{\rm B}}{\mu_0 r_0} + \sqrt{\left(\frac{\pi_{s_n} M_{\rm B}}{\mu_0 r_0} \right)^2 + \left(\frac{\pi_{s_n}}{3 \mu_0 r_0} \right)^3} \right)^{1/3} + \left(\frac{\pi_{s_n} M_{\rm B}}{\mu_0 r_0} - \sqrt{\left(\frac{\pi_{s_n} M_{\rm B}}{\mu_0 r_0} \right)^2 + \left(\frac{\pi_{s_n}}{3 \mu_0 r_0}\right)^3}\right)^{1/3},
\end{eqnarray}
and $\mu_0$ is chosen to guarantee that $\tilde \lambda_{s_1}^\star + \cdots + \tilde \lambda_{s_N}^\star = M_{\rm B}$. In general, these obtained solutions $\tilde \lambda_{s_1}^\star, \cdots, \tilde \lambda_{s_N}^\star$ are not integers, and therefore we need to truncate them to satisfy the integer requirement. Specifically, we can round off $\tilde \lambda_{s_1}^\star, \cdots, \tilde \lambda_{s_N}^\star$ to obtain an integer solution $\lambda_{s_1}^\star, \cdots, \lambda_{s_N}^\star$. However, due to the rounding off operation, the constraint $\lambda_{s_1}^\star + \cdots + \lambda_{s_N}^\star = M_{\rm B}$ may be slightly violated. To tackle this problem, when $\lambda_{s_1}^\star + \cdots + \lambda_{s_N}^\star = M > M_{\rm B}$ and $K = M - M_{\rm B}$, we calculate $d_n = \lambda_{s_n}^\star - \tilde \lambda_{s_n}^\star$, $\forall n = 1, \cdots, N$. If $d_{p_1} > d_{p_2} > \cdots > d_{p_N}$, we let $\lambda_{s_{p_k}}^\star = \lambda_{s_{p_k}}^\star - 1$, $\forall k = 1, \cdots, K$. On the other hand, when $\lambda_{s_1}^\star + \cdots + \lambda_{s_N}^\star = M < M_{\rm B}$ and $K = M_{\rm B} - M$, we compute $d_n = \tilde \lambda_{s_n}^\star - \lambda_{s_n}^\star$, $\forall n = 1, \cdots, N$, and if $d_{p_1} > d_{p_2} > \cdots > d_{p_N}$, we let $\lambda_{s_{p_k}}^\star = \lambda_{s_{p_k}}^\star + 1$, $\forall k = 1, \cdots, K$. The other subproblems can be solved similarly and we summarize the whole procedure in \textbf{Algorithm \ref{KKT}} for clarity.

\begin{algorithm}[!htbp]
\caption{Exploit Karush-Kuhn-Tucker Conditions to Solve \textbf{(P3)}} \label{KKT}
\SetKwInOut{Input}{Input}\SetKwInOut{Output}{Output}
\Input{The total number of training beam pairs $M_{\rm B}$.}
\Output {A suboptimal solution to \textbf{(P3)}.}
Initialization: $\lambda_{s_n} = 1$, $\forall n = 1, \cdots, M_{\rm B}$, and let $\Delta = \bar \Gamma_1^{\rm apx}(\bm \lambda_{M_{\rm B}})$. \\
Let $N = M_{\rm B} - 1$ and ${\cal T}(N) = \pi_{s_1} + \cdots + \pi_{s_N}$. \\
\While {${\cal T}(N) > \Delta$}
{
    Relax the I-NLP subproblem and solve the relaxed problem via using its KKT conditions. The associated solution is expressed as $\tilde \lambda_{s_n}^\star = \max\{1, b_n\}$, where $b_n$ is given by \eqref{LCS-4}, $\forall n = 1, \cdots, N$. \\
    Truncate $\tilde \lambda_{s_1}^\star, \cdots, \tilde \lambda_{s_N}^\star$ to obtain an integer solution $\left\{\lambda_{s_1}^\star, \cdots, \lambda_{s_N}^\star\right\}$, where some modifications might be needed to satisfy the constraint $\sum\nolimits_{n = 1}^N \lambda_{s_n}^\star = M_{\rm B}$.\\[4pt]
    Update the objective value: $\Delta = \bar \Gamma_1^{\rm apx}(\bm \lambda_N^\star)$ if $\bar \Gamma_1^{\rm apx}(\bm \lambda_N^\star) > \Delta$, otherwise, $\Delta$ remains unchanged. \\
    Let $N = N - 1$. \\
}
\end{algorithm}

\textbf{Remark 1}: It is worth mentioning that the proposed beam pair allocation strategy is deduced by assuming accurate knowledge of $\theta^{[\tau - 1]}$ and $\vartheta^{[\tau - 1]}$. If the previous estimates of $\theta^{[\tau - 1]}$ and $\vartheta^{[\tau - 1]}$ are inaccurate, the proposed strategy might worsen the $\tau$-th and the subsequent beam tracking procedures, as demonstrated in Section \ref{NR}. To alleviate the error propagation phenomenon incurred by the proposed beam allocation strategy, when $\big|\xi_{z_n}^{[\tau - 1]} \big|^2 - \big|\xi_{z_m}^{[\tau - 1]}\big|^2 < \Omega$, where $\big|\xi_{z_n}^{[\tau - 1]}\big|^2$ and $\big |\xi_{z_m}^{[\tau - 1]}\big|^2$ are the two largest received signal powers in the $(\tau - 1)$-th beam training period and $\Omega$ is a pre-defined threshold, we employ uniform allocation strategy instead of invoking the proposed allocation strategy in the $\tau$-th beam training period.

\section{Extension to General Non-Orthogonal Codebook Matrices} \label{S-NOBP}
In this section, we consider a more general scenario in which $N_{\rm T} < X_{\rm T}$ and $N_{\rm R} < X_{\rm R}$, such that the two beam codebook matrices ${\bf A}_{\rm T}$ and ${\bf A}_{\rm R}$ are not DFT matrices any longer. In this case, even if the adopted Tx-Rx beam pair is not perfectly aligned with the actual AoA and AoD, the MS can still receive the pilot symbol with a considerable beamforming gain, which is different from the orthogonal case.

\subsection{Power-Based Estimator}
As before, we pick the columns from ${\bf A}_{\rm T}$ and ${\bf A}_{\rm R}$ as the Tx and Rx beams in each beam training period, and when ${\bf f}^{[m]} = {\bf a}_{\rm T}(\bar \vartheta_i)$ and ${\bf w}^{[m]} = {\bf a}_{\rm R}(\bar \theta_k)$ are chosen at the $m$-th symbol duration, the received symbol in \eqref{ATEP-1} becomes
\begin{eqnarray}\label{NOBP-1}
    y_{k, i} = \gamma \alpha {\bf a}^{\rm H}_{\rm R}(\bar \theta_k) {\bf a}_{\rm R}(\bar \theta_{k_1}) {\bf a}^{\rm H}_{\rm T}(\bar \vartheta_{i_1}) {\bf a}_{\rm T}(\bar \vartheta_i) + n^{[m]} = \gamma \nu_{k, k_1} {\tilde \nu}_{i, i_1} \alpha + n^{[m]},
\end{eqnarray}
where ${\tilde \nu}_{i, i_1}$ and $\nu_{k, k_1}$ are respectively expressed as
\begin{eqnarray}
    && {\tilde \nu}_{i, i_1} ~~=~ {\bf a}_{\rm T}^{\rm H} (\bar \vartheta_{i_1}) {\bf a}_{\rm T}(\bar \vartheta_i), \label{NOBP-2A} \\
    && \nu_{k, k_1} ~=~ {\bf a}_{\rm R}^{\rm H} (\bar \theta_k) {\bf a}_{\rm R}(\bar \theta_{k_1}). \label{NOBP-2B}
\end{eqnarray}
It is easy to see that when $i = i_1$ and $k = k_1$, $\tilde \nu_{i, i_1} = 1$ and $\nu_{k, k_1} = 1$, and \eqref{NOBP-1} reduces to \eqref{ATEP-1}. Following the previous description, the beam pairs ${\cal B}_{z_1}, \cdots, {\cal B}_{z_N}$ with repetition times $\lambda_{z_1}, \cdots, \lambda_{z_N}$ are used in the current beam training period, where $z_p = (a_p \bullet c_p)_{X_{\rm R}}$, $p = 1, \cdots, N$. Based on this assumption, the distribution of $\xi_{z_p}$, which is first defined in \eqref{ATEP-2}, is given by
\begin{equation}\label{NOBP-4}
    \xi_{z_p} = \sum\limits_{m = 1}^{\lambda_{z_p}} {\bf y}_{z_p}[m] \sim {\cal CN}\Big(\gamma \alpha \nu_{a_p, k_1} \tilde \nu_{c_p, i_1} \lambda_{z_p}, \sigma_0^2 \lambda_{z_p} \Big).
\end{equation}

The successful estimation probability shown in \eqref{ATEP-4} now becomes
\begin{eqnarray}\label{NOBP-5}
    \Gamma_{z_n, |\alpha|^2} = \int\limits_0^\infty h \bigg(u; \lambda_{z_n}, |\alpha|^2 \bigg) \prod\limits_{m = 1, m \ne n}^N \left(1 - Q_1 \left(\sqrt{\frac{2 \lambda_{z_m} \gamma^2 |\nu_{a_m, k_1} \tilde \nu_{c_m, i_1} \alpha|^2} {\sigma_0^2}}, \sqrt{\frac{2 u} {\lambda_{z_m} \sigma_0^2}} \right) \right) \text{d} u,
\end{eqnarray}
where the non-central chi-squared distribution $h(u; \lambda_{z_n}, |\alpha|^2)$ is already given by \eqref{ATEP-5} and $Q_1$ is the first-order Marcum $Q$-function. Furthermore, by integrating $\Gamma_{z_n, |\alpha|^2}$ over the exponential distribution of $|\alpha|^2$, we obtain
\begin{eqnarray}\label{NOBP-6}
    \Gamma_{z_n} & = & \int\limits_0^\infty \int\limits_0^\infty h \bigg(u; \lambda_{z_n}, |\alpha|^2\bigg) \prod\limits_{m = 1, m \ne n}^N \left(1 - Q_1 \left(\sqrt{\frac{2 \lambda_{z_m} \gamma^2 |\nu_{a_m, k_1} \tilde \nu_{c_m, i_1} \alpha|^2} {\sigma_0^2}}, \sqrt{\frac{2 u}{\lambda_{z_m} \sigma_0^2}} \right) \right) \nonumber \\
    & \times & \frac{1}{\sigma_\alpha^2} \exp\left(-\frac{|\alpha|^2} {\sigma_\alpha^2}\right) \text{d} u ~\text{d} |\alpha|^2.
\end{eqnarray}
It is challenging to derive a closed-form expression for $\Gamma_{z_n}$ in \eqref{NOBP-6}, which is therefore calculated via the numerical method. Finally, similar to \eqref{ATEP-8}, the ASTP can be given by
\begin{equation}\label{NOBP-7}
    \bar \Gamma_2 = \sum\limits_{n = 1}^N \pi_{z_n} \times \Gamma_{z_n}.
\end{equation}

\subsection{OMP-Based Estimator} \label{S-OMP}
In general, the power-based estimator performs poorly due to the non-negligible inter-beam interference, which actually can be used to improve the beam tracking performance. To show this, we first rewrite the channel in \eqref{SM-3} as
\begin{equation}
    {\bf H} = \sqrt{N_{\rm T} N_{\rm R}} \alpha {\bf A}_{\rm R} {\bf V} {\bf A}^{\rm H}_{\rm T},
\end{equation}
where $\bf V$ is termed the beamspace channel representation of $\bf H$ in some literature \cite{Sayeed}. Based on the on-grid assumption of the AoAs and AoDs, the matrix $\bf V$ is sparse with only $L$ nonzero elements, e.g., ${\bf V}[k_1, i_1]$ in current transmission block.

By introducing the beamspace channel $\bf V$, we can transform the AoA and AoD estimation problem into a CS problem. To be more specific, we rewrite \eqref{ATEP-1} as
\begin{eqnarray}\label{NOBP-8}
    y_{k, i} & = & ({\bf w}^{[m]})^{\rm H} {\bf H} {\bf f}^{[m]} x^{[m]} + n^{[m]} \\
    \hspace{-0.1cm} & = & \gamma \alpha ({\bf w}^{[m]})^{\rm H} {\bf A}_{\rm R} {\bf V} {\bf A}_{\rm T}^{\rm H} {\bf f}^{[m]} + n^{[m]} \nonumber \\
    & \overset{(a)}{=} & \gamma \alpha [({\bf A}_{\rm T}^{\rm H} {\bf f}^{[m]})^{\rm T} \otimes (({\bf w}^{[m]})^{\rm H} {\bf A}_{\rm R})] {\bf v} + n^{[m]} \nonumber \\
    & \overset{(b)}{=} & \underbrace{\gamma \alpha ([\tilde \nu_{i, 1}, \cdots, \tilde \nu_{i, X_{\rm T}}] \otimes [\nu_{k, 1}, \cdots, \nu_{k, X_{\rm R}}])}_{{\bf a}_n^{\rm T}} {\bf v} + n^{[m]}, \nonumber
\end{eqnarray}
where ${\bf v} = \text{vec}({\bf V})$. In \eqref{NOBP-8}, we have applied the property $\text{vec}({\bf A B C}) = ({\bf C}^{\rm T} \otimes {\bf A}) \text{vec}({\bf B})$ in (a) and (b) is due to \eqref{NOBP-2A} and \eqref{NOBP-2B}. Recall that ${\cal B}_{z_1}, \cdots, {\cal B}_{z_N}$ are used, and if we collect their corresponding observations into a vector, denoted by
\begin{equation*}\label{NOBP-9}
    {\bf y} = \Big[{\bf y}_{z_1}[1], \cdots, {\bf y}_{z_1}[\lambda_{z_1}], \cdots, {\bf y}_{z_N}[1], \cdots, {\bf y}_{z_N}[\lambda_{z_N}] \Big]^{\rm T},
\end{equation*}
we can obtain a CS problem, given by
\begin{equation}\label{NOBP-10}
    {\bf y} = {\bf A} {\bf v} + {\bf n},
\end{equation}
where ${\bf n} = \Big[n^{[1]}, \cdots, n^{[M_{\rm B}]}\Big]^{\rm T}$. The sensing matrix $\bf A$ is written as
\begin{equation}\label{NOBP-11}
{\bf A} = \Big[\underbrace{{\bf a}_{z_1}, \cdots, {\bf a}_{z_1}}_{\lambda_{z_1}}, \cdots, \underbrace{{\bf a}_{z_N}, \cdots, {\bf a}_{z_N}}_{\lambda_{z_N}}\Big]^{\rm T},
\end{equation}
where ${\bf a}_{z_n}$ is shown in \eqref{NOBP-8} through replacing $k$ and $i$ by $a_n$ and $c_n$, respectively, $n = 1, \cdots, N$.

The MS can use the OMP algorithm \cite{OMP, OMP2} to estimate the nonzero element in $\bf v$, which is also termed the support of $\bf v$ \cite{Duan-Tracking}. By adopting the OMP algorithm, the estimated support of $\bf v$ is given by
\begin{equation}\label{NOBP-12}
    \text{supp}({\bf v}) = \argmax_{1 \le k \le X} ~|{\bf A}^{\rm H} {\bf y}|^2 = \argmax_{1 \le k \le X} ~|\bm \xi|^2,
\end{equation}
where $\bm \xi \triangleq {\bf A}^{\rm H} {\bf y}$. This OMP-based estimator can be viewed as an improvement of the previous power-based estimator, and the successful estimation probability conditioned on $|\alpha|^2$ is given by
\begin{equation}\label{NOBP-13}
    \Gamma_{n_1, |\alpha|^2} = \Pr\Bigg(\bigcap\limits_{n = 1, n \ne n_1}^{X} \big|\bm \xi[n_1]\big|^2 > \big|\bm \xi[n]\big|^2 ~\Big |~ \alpha \Bigg),
\end{equation}
where $n_1 = (k_1 \bullet i_1)_{X_{\rm R}}$. Moreover, we need to integrate \eqref{NOBP-13} over the exponential distribution of $|\alpha|^2$, given by
\begin{equation}\label{NOBP-13A}
    \Gamma_{n_1} = \int\limits_0^\infty \Gamma_{n_1, |\alpha|^2} \times  \frac{1}{\sigma_\alpha^2} \exp\left(-\frac{|\alpha|^2} {\sigma_\alpha^2} \right) \text{d} |\alpha|^2,
\end{equation}
and the ASTP of this OMP-based estimator is expressed as
\begin{equation}\label{NOBP-13B}
    \bar \Gamma_3 = \sum\limits_{n_1 = 1}^X \pi_{n_1} \times \Gamma_{n_1}.
\end{equation}

\subsection{Problem Formulation}
In this section, we replace $\bar \Gamma_1$ by $\bar \Gamma_3$ and construct a new optimization problem, given by
\begin{eqnarray}
    \textbf{(P5)} & \max\limits_{\bm \lambda} & \bar \Gamma_3(\bm \lambda) \label{P5-A} \\
    & {\rm s.t.} & \eqref{P1-B}, \eqref{P1-C}, \text{and}~ \eqref{P1-D}.
\end{eqnarray}
However, different from $\bar \Gamma_1$ in \eqref{ATEP-8}, $\bar \Gamma_3$ has no closed-form expression, and therefore we derive a closed-form lower bound for $\Gamma_{n_1}$, which is shown in \textbf{Lemma \ref{lemma4}}.

\begin{lemma}\label{lemma4}
    If the Tx-Rx beam pairs ${\cal B}_{z_1}, \cdots, {\cal B}_{z_N}$ are used in the beam training period, where $z_m = (a_m \bullet c_m)_{X_{\rm R}}$, $\forall m = 1, \cdots, N$, and their repetition times are respectively denoted by $\lambda_{z_1}, \cdots, \lambda_{z_N}$, the lower bound of $~\Gamma_{n_1, |\alpha|^2}$ is expressed as
    \begin{eqnarray}\label{NOBP-14}
        \Gamma_{n_1, |\alpha|^2}^{\rm lb} & = & 1 - \sum\limits_{n = 1, n \ne n_1}^X \Bigg[ Q_1\left(\sqrt{A_{n, n_1} |\alpha|^2}, \sqrt{B_{n, n_1} |\alpha|^2}\right) - \frac{\sum\nolimits_{m = 1}^N \lambda_{z_m} |\nu_{a_m, k_1} \tilde \nu_{c_m, i_1}|^2}{\sum\nolimits_{m = 1}^N \lambda_{z_m} \Big(|\nu_{a_m, k_1} \tilde \nu_{c_m, i_1}|^2 + |\nu_{a_m, k} \tilde \nu_{c_m, i}|^2\Big)} \nonumber \\
        & \times & \exp\left(-\frac{A_{n, n_1} |\alpha|^2 + B_{n, n_1} |\alpha|^2}{2}\right) I_0 \left(\sqrt{A_{n, n_1} B_{n, n_1}}|\alpha|^2\right)\Bigg],
    \end{eqnarray}
    where $n = (k \bullet i)_{X_{\rm R}}$ and $n_1 = (k_1 \bullet i_1)_{X_{\rm R}}$. Accordingly, the lower bound of $~\Gamma_{n_1}$ is written as
    \begin{eqnarray}\label{NOBP-15}
        \Gamma_{n_1}^{\rm lb} & = & 1 - \sum\limits_{n = 1, n \ne n_1}^X \left(\frac{1}{2} - \frac{(B_{n, n_1} - A_{n, n_1}) \sigma_\alpha^2 - 2}{4 \sqrt{1 + (A_{n, n_1} + B_{n, n_1})\sigma_\alpha^2 + \sigma_\alpha^4 (A_{n, n_1} - B_{n, n_1})^2 \big / 4}}\right) \nonumber \\
        & + & \sum\limits_{n = 1, n \ne n_1}^X \Bigg(\frac{\sum\nolimits_{m = 1}^N \lambda_{z_m} |\nu_{a_m, k_1} \tilde \nu_{c_m, i_1}|^2}{\sum\nolimits_{m = 1}^N \lambda_{z_m} \Big[|\nu_{a_m, k_1} \tilde \nu_{c_m, i_1}|^2 + |\nu_{a_m, k} \tilde \nu_{c_m, i}|^2\Big]} \nonumber \\
        & \times & \frac{1}{\sqrt{1 + (A_{n, n_1} + B_{n, n_1})\sigma_\alpha^2 + \sigma_\alpha^4 (A_{n, n_1} - B_{n, n_1})^2 \big / 4}} \Bigg),
    \end{eqnarray}
    where $A_{n, n_1}$ and $B_{n, n_1}$ are respectively given by
    \begin{equation*}
        A_{n, n_1} = \frac{2 \gamma^2 \left|\sum\nolimits_{m = 1}^N \lambda_{z_m} \nu_{a_m, k}^{\ast} \tilde \nu_{c_m, i}^{\ast} \nu_{a_m, k_1} \tilde \nu_{c_m, i_1} \right|^2}{\sigma_0^2 \sum\nolimits_{m = 1}^N \lambda_{z_m} \Big(|\nu_{a_m, k_1} \tilde \nu_{c_m, i_1}|^2 + |\nu_{a_m, k} \tilde \nu_{c_m, i}|^2\Big)},
    \end{equation*}
    \begin{equation*}
        B_{n, n_1} = \frac{2 \gamma^2 \left(\sum\nolimits_{m = 1}^N \lambda_{z_m}|\nu_{a_m, k_1} \tilde \nu_{c_m, i_1}|^2 \right)^2}{\sigma_0^2 \sum\nolimits_{m = 1}^N \lambda_{z_m} \Big(|\nu_{a_m, k_1} \tilde \nu_{c_m, i_1}|^2 + |\nu_{a_m, k} \tilde \nu_{c_m, i}|^2\Big)}.
    \end{equation*}
\end{lemma}
\begin{IEEEproof}
    Refer to Appendix \ref{lemma4-proof}.
\end{IEEEproof}

According to \textbf{Lemma \ref{lemma4}}, the lower bound for the ASTP when using the OMP-based estimator can be written as
\begin{equation}\label{NOBP-18}
    \bar \Gamma_3 = \sum\limits_{n_1 = 1}^X \pi_{n_1} \times \Gamma_{n_1} > \sum\limits_{n_1 = 1}^X \pi_{n_1} \times \Gamma_{n_1}^{\rm lb} \triangleq \bar \Gamma_3^{\rm lb}.
\end{equation}

Since we cannot directly optimize $\bar \Gamma_3$ in \eqref{P5-A}, we use $\bar \Gamma_3^{\rm lb}$ as the new objective function, and the associated optimization problem becomes
\begin{eqnarray}
    \textbf{(P6)} & \max\limits_{\bm \lambda} & \bar \Gamma_3^{\rm lb}(\bm \lambda) \label{P6-A} \\
    & {\rm s.t.} & \eqref{P1-B}, \eqref{P1-C}, \text{and}~ \eqref{P1-D}.
\end{eqnarray}
Considering the complicated expression for $\bar \Gamma_3^{\rm lb}$, it is still very challenging to solve \textbf{(P6)} analytically. Therefore, when $M_{\rm B}$ is small, we propose to directly search for the $M_{\rm B}$ Tx-Rx training beam pairs from the two beam codebook matrices ${\bf A}_{\rm T}$ and ${\bf A}_{\rm R}$ that can maximize \eqref{P6-A}. On the other hand, when $M_{\rm B}$ is large, this exhaustive search method is prohibitive since there are $X^{M_{\rm B}}$ possible solutions. In this situation, we can exploit a heuristic algorithm, such as the differential evolution \cite{DE-Survey}, to obtain a promising solution to \textbf{(P6)}, whose details are omitted for brevity.

\textbf{Remark 2}: Both the power-based estimator and the OMP-based estimator can be readily extended to the multi-path scenario as in \cite{RH-MSHC, MAP}. To be more specific, we can estimate the $L \ge 2$ paths in an iterative fashion, and only one path is estimated via using the power-based estimator or the OMP-based estimator at each iteration. Following the idea of successive interference cancelation, the contribution of the paths that have been estimated in the previous iterations is subtracted from the received sequence before finding new paths. In addition, since jointly optimizing the $M_{\rm B}$ Tx-Rx beam pairs is very challenging when multiple paths exist, we optimize the $M_{\rm B} / L$ training beams pairs for each of the $L$ paths via solving an optimization problem similar to \textbf{(P3)} or \textbf{(P6)} separately. Though such a per-path beam allocation strategy seems to be trivial, it can still achieve a favorable beam tracking performance.

\section{Numerical Results} \label{NR}
In this section, we provide numerical results to evaluate the tracking performance of the proposed beam pair allocation strategy for time-varying mmWave MIMO systems. The average tracking error probability (ATEP) is mainly used as the performance metric, which is expressed as $1 - \text{ASTP}$. The geometric channel model in \eqref{SM-3} is adopted with $L = 1$ and $\sigma_\alpha^2 = 1$. The SNR is defined as $P / \sigma_0^2$. To guarantee a favorable angular resolution, we require that $X_{\rm T} = X_{\rm R} = 64$. In addition, one transmission frame consists of $T = 10$ transmission blocks \cite{Duan-Tracking, POMDP} and the temporal variations of AoA and AoD among these transmission blocks are assumed to follow two discrete Markov processes, described by \eqref{SM-5A} and \eqref{SM-5B}, respectively. Moreover, the first transmission block is assumed to have exact AoA and AoD knowledge by using the traditional channel estimation algorithms such as \cite{RH-MSHC}. For the remaining 9 blocks, we exploit the previously estimated AoA and AoD to conduct the current beam tracking procedure. Several benchmark methods are also introduced for comparison, which are presented as follows:

\begin{enumerate}
\item Proportional Allocation: The $M_{\rm B}$ pilot symbol durations are distributed to ${\cal B}$ in proportion to their associated transition probabilities;
\item Uniform Allocation: The $M_{\rm B}$ pilot symbol durations are uniformly distributed to ${\cal B}_{s_1}, \cdots, {\cal B}_{s_{M_{\rm B}}}$;
\item Proposed Allocation ES: The $M_{\rm B}$ pilot symbol durations are distributed to ${\cal B}$ via solving \textbf{(P2)} or \textbf{(P6)} with an exhaustive search method;
\item Proposed Allocation BB: The $M_{\rm B}$ pilot symbol durations are distributed to ${\cal B}$ via solving \textbf{(P3)} with \textbf{Algorithm 1};
\item Proposed Allocation KKT: The $M_{\rm B}$ pilot symbol durations are distributed to ${\cal B}$ via solving \textbf{(P4)} with \textbf{Algorithm 2};
\item ML-Based Estimator: The time-varying AoA and AoD of a single-path channel are estimated based on the ML criterion \cite{ML};
\item POMDP Framework: The time-varying AoA and AoD are estimated based on the belief states of the formulated POMDP in \cite{POMDP}.
\end{enumerate}

\begin{figure}
\begin{minipage}{.49\textwidth}
\centering
\includegraphics[width = 8.8cm]{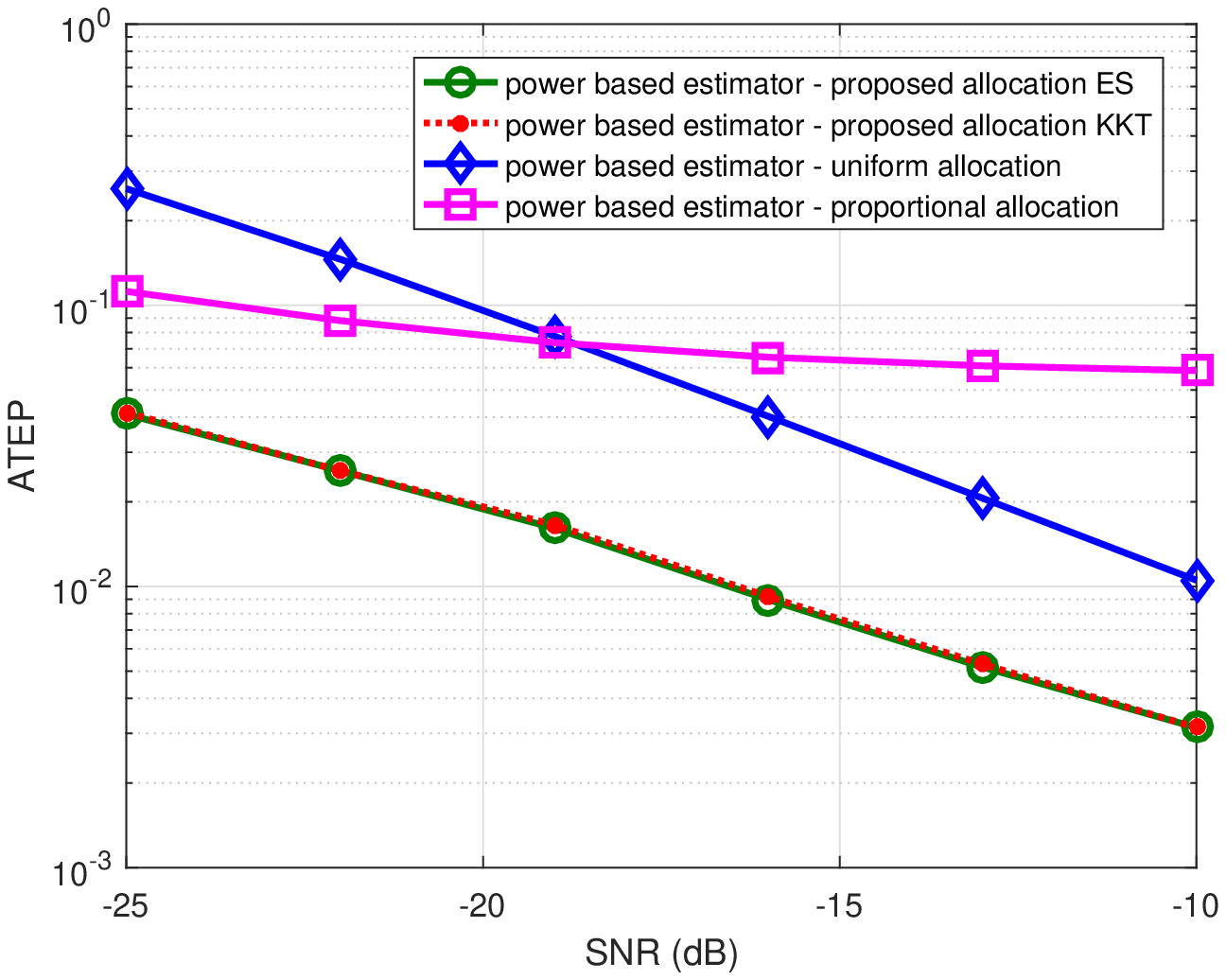}
\caption{The ATEP with respect to the training SNR. $N_{\rm T} = 64$, $N_{\rm R} = 64$, $\beta = 0.1$, $\tilde \beta = 0.1$, and $M_{\rm B} = 40$.} \label{BER-SNR2}
\end{minipage}
\begin{minipage}{.49\textwidth}
\centering
\includegraphics[width = 8.8cm]{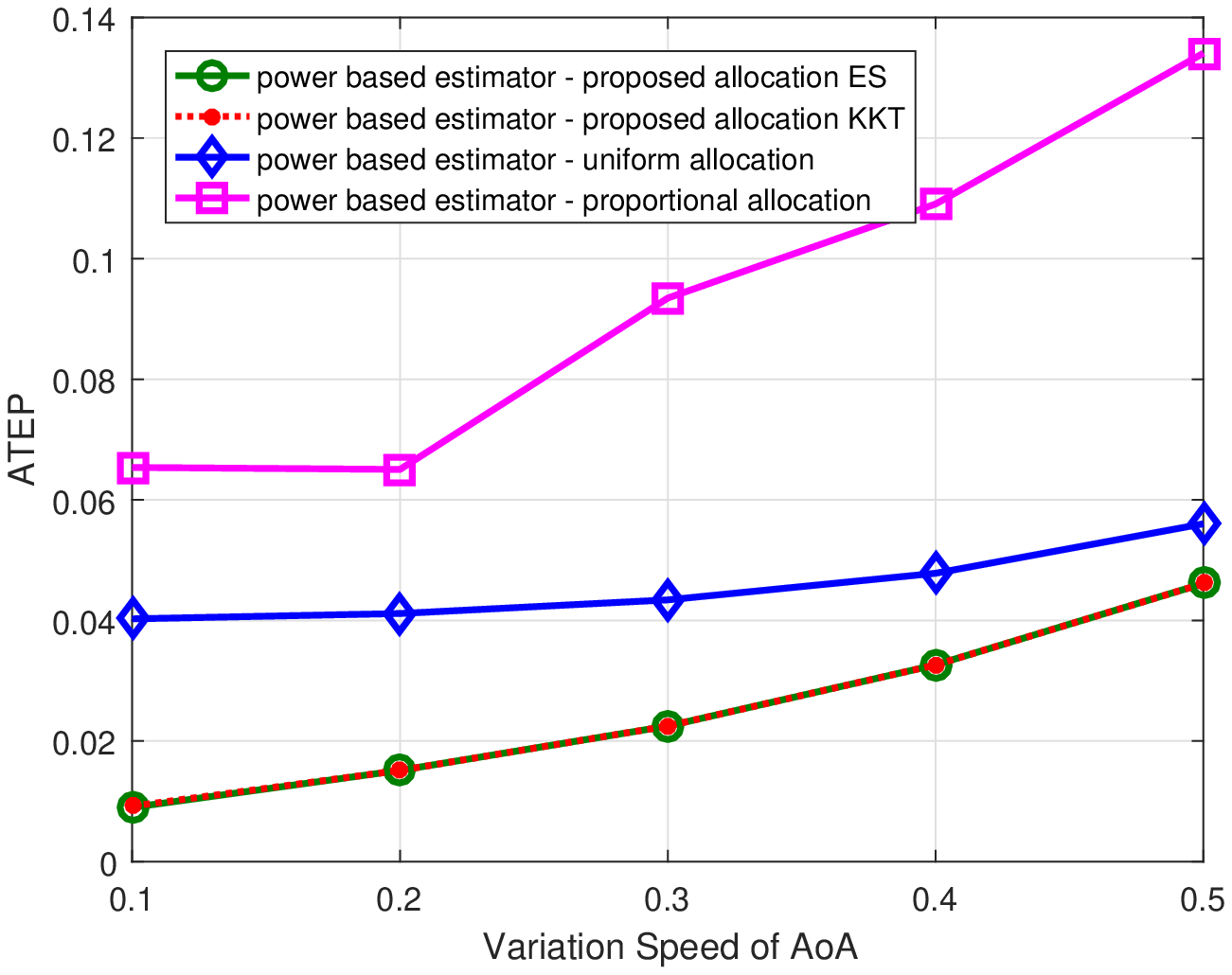}
\caption{The ATEP with respect to the variation speed of AoA. $N_{\rm T} = N_{\rm R} = 64$, $\tilde \beta = 0.1$, $M_{\rm B} = 40$, and SNR = -16dB.} \label{BER-BETA2}
\end{minipage}
\end{figure}

\subsection{Special Scenario: Orthogonal Tx-Rx Beam Pairs}
The ATEP with respect to the training SNR is provided in Fig. \ref{BER-SNR2}, where we let $N_{\rm T} = X_{\rm T}$ and $N_{\rm R} = X_{\rm R}$. In this scenario, we see that the proportional allocation strategy outperforms the uniform allocation strategy in the low SNR region, whereas it is inferior to the latter when the training SNR is high. The proposed beam allocation strategy performs better than the two benchmarks at the whole SNR range. In addition, we note that the solution obtained by solving the $M_{\rm B}$ relaxed subproblems with KKT conditions in \textbf{(P4)} can achieve almost the same ATEP as that obtained via the ES method. It is also worth mentioning that the ATEP curve of the iterative N-BB algorithm is not presented in Fig.\ref{BER-SNR2} for clarity, since it converges to the curves of the ES and KKT-based methods.

The ATEP with respect to the AoA's variation speed $\beta$ is shown in Fig. \ref{BER-BETA2}, where we can see that the proposed beam allocation strategy still performs better than the other three strategies, though the ATEPs of the four aforementioned beam allocation strategies all deteriorate when $\beta$ increases from 0.1 to 0.5. Moreover, the ATEP gap between the proposed beam allocation strategy and the uniform allocation strategy becomes narrower and narrower. The reason can be explained by observing \eqref{LCS-4}, which demonstrates that the repetition times of each training beam pair is directly determined by its associated transition probability. Since the transition probability for each of the potential directions becomes more uniform when $\beta$ is large, according to \eqref{LCS-4}, the proposed beam allocation strategy is asymptotically close to the uniform allocation strategy.

\begin{figure}
\begin{minipage}{.49\textwidth}
\centering
\includegraphics[width = 8.8cm]{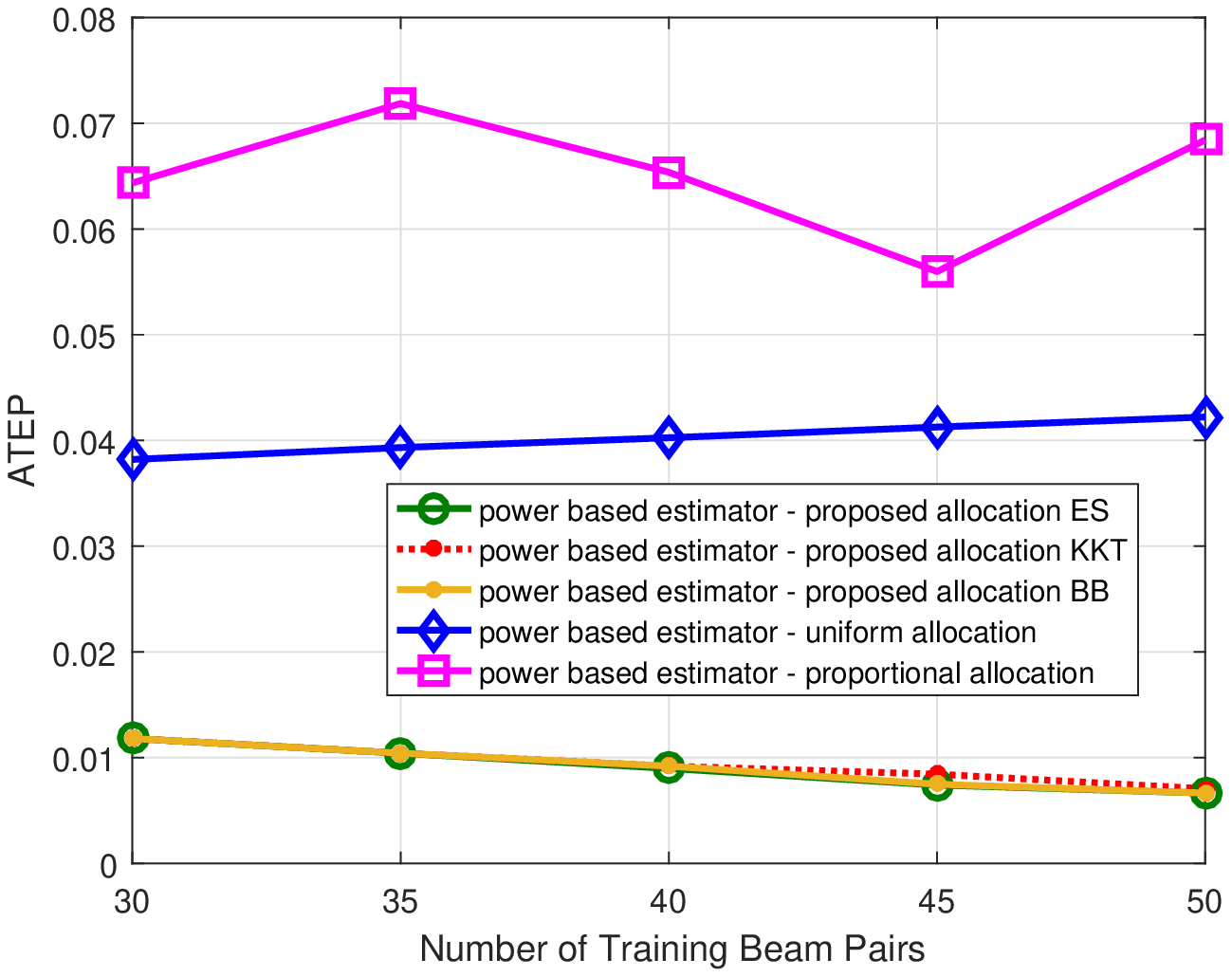}
\caption{The ATEP with respect to the number of training beam pairs. $N_{\rm T} = N_{\rm R} = 64$, $\beta = \tilde \beta = 0.1$, and SNR = -16dB.} \label{BER-NUM2}
\end{minipage}
\begin{minipage}{.49\textwidth}
\centering
\includegraphics[width = 8.8cm]{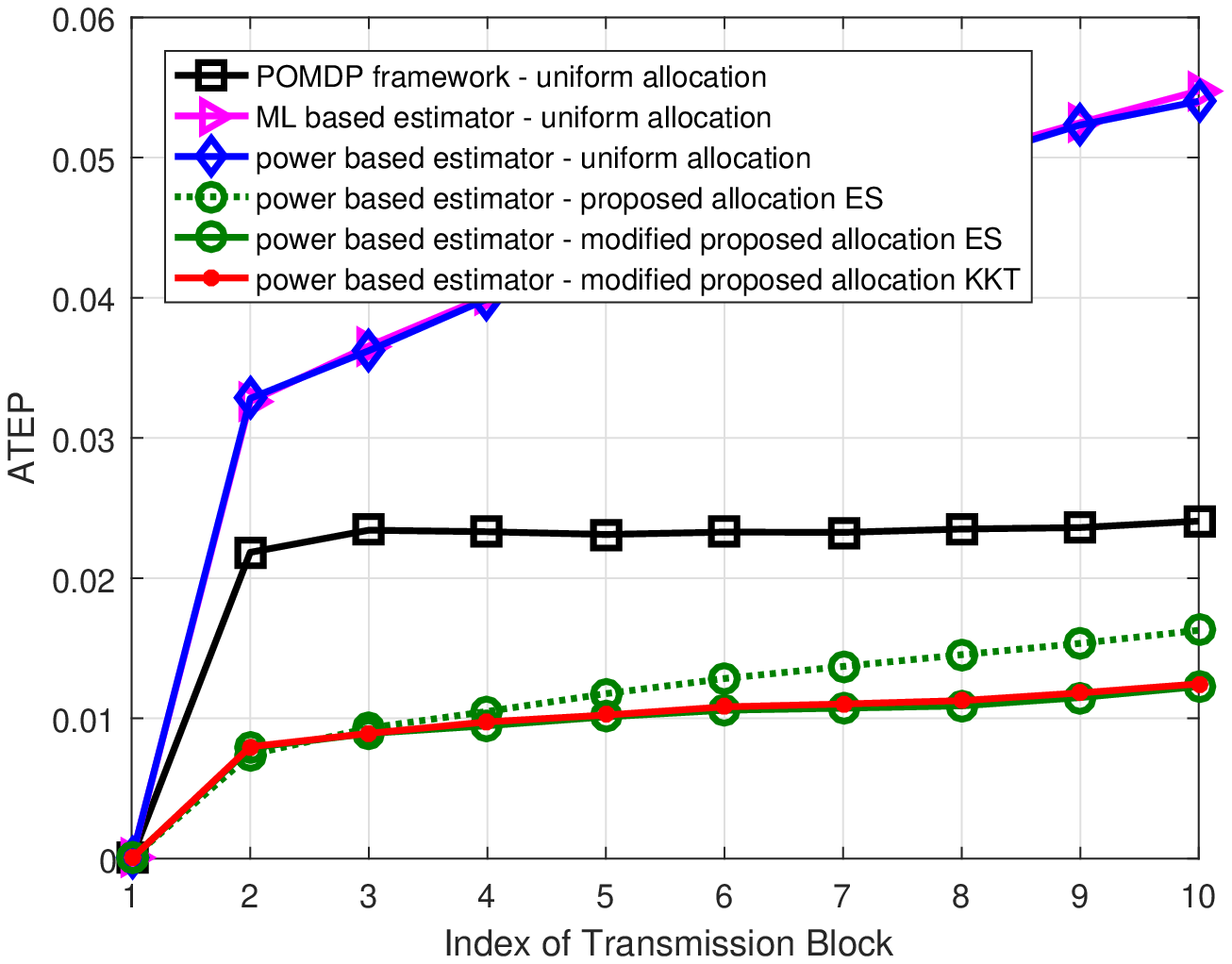}
\caption{The ATEP with respect to the index of transmission block. $N_{\rm T} = N_{\rm R} = 64$, $\beta = \tilde \beta = 0.1$, $M_{\rm B} = 40$, and SNR = -16dB.} \label{BER-TIME2}
\end{minipage}
\end{figure}

The ATEP with respect to the total number of training beam pairs is presented in Fig. \ref{BER-NUM2}, where we can see that the ATEP of the proposed allocation strategy decreases gradually when $M_{\rm B}$ increases from 30 to 50. The ATEP of the KKT-based algorithm is very close to that of the iterative N-BB algorithm, which achieves almost the same performance as the ES method, demonstrating the validity of \textbf{Algorithm 1} and \textbf{Algorithm 2}. Moreover, we observe that the ATEP of the uniform beam allocation strategy deteriorates slightly as $M_{\rm B}$ increases. According to \eqref{ATEP-7} and \eqref{ATEP-8}, though more directions can be measured when we increase $M_{\rm B}$, the interference in \eqref{ATEP-7} also increases such that $\Gamma_{z_n}$ will decrease, and consequently the ATEP of the the uniform beam allocation strategy increases.

The ATEP with respect to the index of transmission block is shown in Fig. \ref{BER-TIME2}, in which we observe that the performance of the POMDP framework is quite robust with the process of beam tracking. The ATEPs of the ML-based estimator and the proposed power-based estimator degrade quickly compared to that of the POMDP framework. However, by exploiting the modified beam allocation strategy depicted in \textbf{Remark 1}, it is seen that the ATEP performance of the power-based estimator can be significantly improved, where we set the threshold $\Omega = 5$ in the simulations.

\begin{figure}
\centering
\subfigure[]
{\includegraphics[width = 8.8cm]{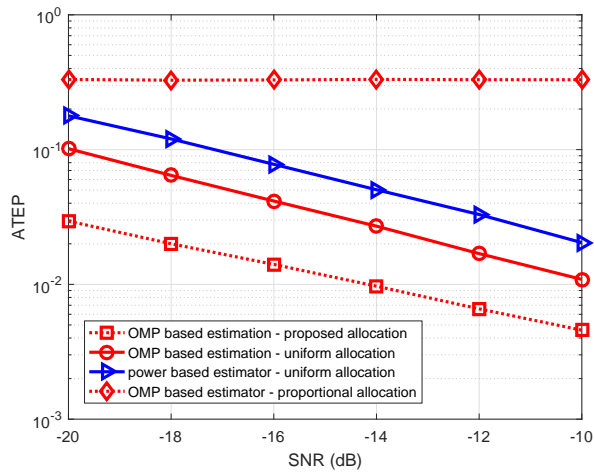} \label{BER-SNR1}}
\subfigure[]
{\includegraphics[width = 8.8cm]{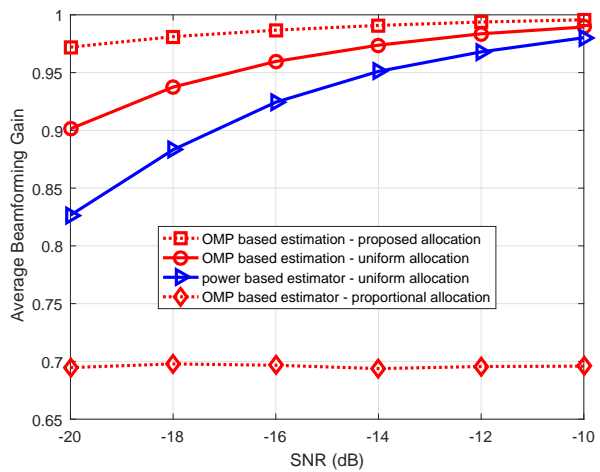} \label{RATE-SNR1}}
\caption{The ATEP (a) and average beamforming gain (b) with respect to SNR. $N_{\rm T} = 48$, $N_{\rm R} = 48$, $\beta = 0.1$, $\tilde \beta = 0.1$, and $M_{\rm B} = 40$.}
\end{figure}

\subsection{General Scenario}
The ATEP with respect to the SNR when $N_{\rm T} < X_{\rm T}$ and $N_{\rm R} < X_{\rm R}$ is shown in Fig. \ref{BER-SNR1}, where we can immediately see that the proportional allocation strategy cannot work in the whole SNR region. The power-based estimator in \eqref{NOBP-5} is inferior to the OMP-based estimator in \eqref{NOBP-12} which exploits the inter-beam interference to estimate the AoA and AoD. In addition, we also observe that the proposed beam allocation strategy performs much better than the uniform and proportional allocation strategies. Moreover, since two adjacent beams are overlapped in the angular space when $N_{\rm T} < X_{\rm T}$ and $N_{\rm R} < X_{\rm R}$, the BS can still transmit information to the MS with a considerable beamforming gain even if the estimates of the true AoA and AoD are inaccurate. Therefore, we also provide the average beamforming gains\footnote{The average beamforming gain is expressed as ${\mathbb E}\big[|{\bf a}^{\rm H}_{\rm R} \big(\hat \theta^{[\tau]}\big) {\bf a}_{\rm R} \big(\theta^{[\tau]}\big) {\bf a}^{\rm H}_{\rm T} \big(\vartheta^{[\tau]}\big) {\bf a}_{\rm T} (\hat \vartheta^{[\tau]}) |^2\big]$, where $\hat \theta^{[\tau]}$ and $\hat \vartheta^{[\tau]}$ represent the estimated AoA and AoD in the $\tau$-th beam training period.} of the four aforementioned strategies in Fig. \ref{RATE-SNR1}, in which we can observe that the proposed beam allocation strategy is still significantly superior to the other three strategies, especially at the low SNR regime.

\begin{figure}
\begin{minipage}{.49\textwidth}
\centering
\includegraphics[width = 8.8cm]{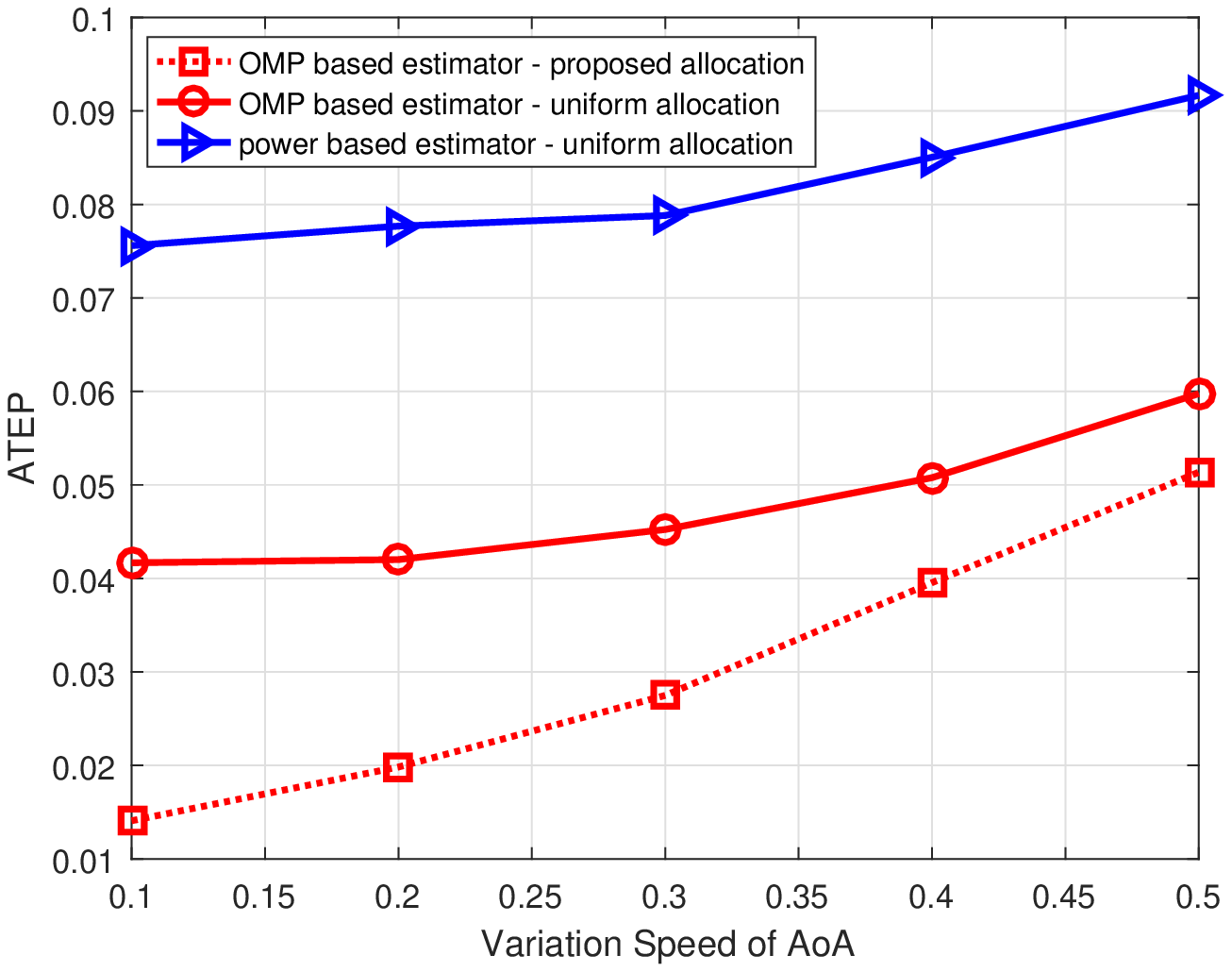}
\caption{The ATEP with respect to the variation speed of AoA. $N_{\rm T} = 48$, $N_{\rm R} = 48$, $\tilde \beta = 0.1$, $M_{\rm B} = 40$, and SNR = -16dB.} \label{BER-BETA1}
\end{minipage}
\begin{minipage}{.49\textwidth}
\centering
\includegraphics[width = 8.8cm]{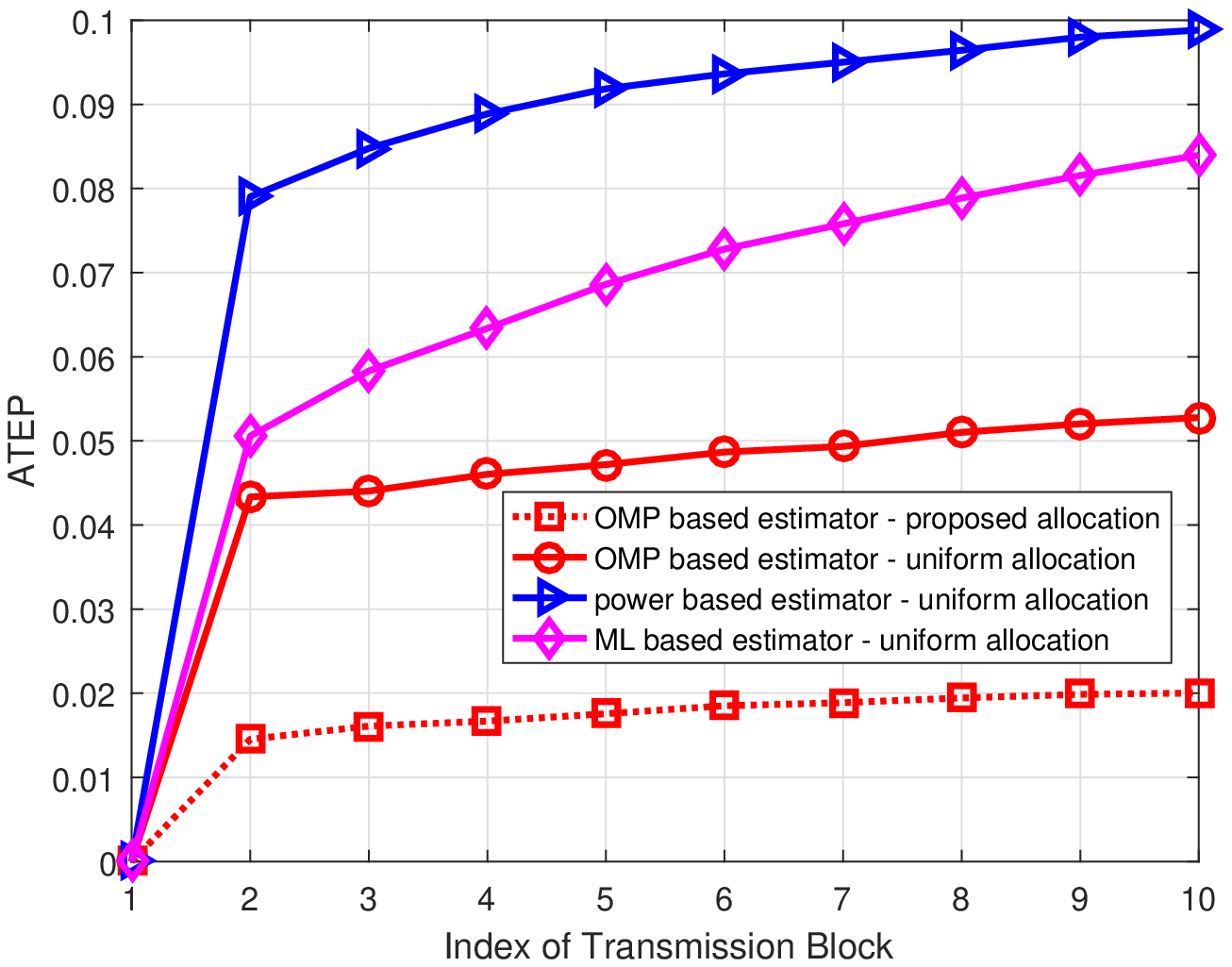}
\caption{The ATEP with respect to the index of transmission block. $N_{\rm T} = N_{\rm R} = 48$, $\beta = \tilde \beta = 0.1$, $M_{\rm B} = 40$, and SNR = -16dB.} \label{BER-TIME1}
\end{minipage}
\end{figure}

The ATEP with respect to $\beta$ is depicted in Fig. \ref{BER-BETA1}, where we fix $\tilde \beta$ at 0.1. Similar to Fig. \ref{BER-BETA2}, while both the ATEPs of the OMP-based estimator with the proposed beam allocation strategy and uniform allocation strategy deteriorate when $\beta$ increases from 0.1 to 0.5, the former strategy still performs better than the latter. Moreover, as $\beta$ increases, we see that the ATEP gap between the two strategies also becomes narrower.

In Fig. \ref{BER-TIME1}, the ATEPs of the 10 transmission blocks are presented, where the AoA and AoD keep changing from one transmission block to another. It is seen that the beam tracking performance of the OMP-based estimator is significantly superior to that of the power-based estimator due to the inter-beam interference. The ML-based estimator performs better than the power-based estimator, whereas it is inferior to the OMP-based estimator. Moreover, we also observe that the ATEP of the OMP-based estimator with the proposed allocation strategy is much better than those of the benchmarks.

\section{Conclusions} \label{CN}
In this paper, we have proposed a new beam pair allocation strategy for mmWave MIMO tracking systems, which enables one Tx-Rx beam pair to be used repeatedly to improve the received signal power at that direction. We have firstly considered a special scenario in which $N_{\rm T} = X_{\rm T}$ and $N_{\rm R} = X_{\rm R}$. In this case, the Tx-Rx beam pairs are orthogonal with each other and the training beam pair sequence design problem can be approximately tackled by solving a set of concave I-NLPs. The obtained closed-form solution shows that the repetition times of each Tx-Rx beam pair is directly determined by its associated transition probability, and one beam pair with a higher transition probability should be used more times than those with lower transition probabilities. In the case of $N_{\rm T} < X_{\rm T}$ and $N_{\rm R} < X_{\rm R}$, we have derived a closed-form lower bound for the ASTP when the OMP-based estimator is used to track the time-varying AoA and AoD, based on which a favorable beam pair allocation strategy is obtained. Our numerical results have validated the superiority of the proposed allocation strategy over the existing methods.

\begin{appendices}
\section{}\label{theorem-proof}
In order to prove \textbf{Theorem 1}, we first show that $\Gamma_{z_n}$ is an increasing function with respect to \mbox{$\lambda_{z_n}$, and} a decreasing function with respect to $\lambda_{z_m}$, $\forall z_m \ne z_n$. To verify this conclusion, we relax the integer variable $\lambda_{z_n}$ to a real variable $\tilde \lambda_{z_n}$, and derive the partial derivative of $\Gamma_{z_n}$ with respect to $\tilde \lambda_{z_n}$ as
\begin{eqnarray}\label{PF-1}
    \frac{\partial \Gamma_{z_n}}{\partial \tilde \lambda_{z_n}}
    & = & \frac {h_2(\tilde \lambda_{z_n})}{h_1^2(\tilde \lambda_{z_n})} ~\int\limits_0^\infty \exp \left(-\frac{u}{h_1(\tilde \lambda_{z_n})} \right) \left(\frac{u}{h_1(\tilde \lambda_{z_n})} - 1 \right) ~\prod\limits_{m = 1, m \ne n}^N \left(1 - \exp \left(-\frac{u}{\lambda_{z_m} \sigma_0^2} \right) \right) \text{d} u \\
    & = & \frac{h_2(\tilde \lambda_{z_n})}{h_1^2(\tilde \lambda_{z_n})} \int\limits_0^\infty \exp\left(-\frac{u} {h_1(\tilde \lambda_{z_n})} \right) \sum\limits_{m \ne n} \frac{u}{\lambda_{z_m} \sigma_0^2}\exp\left(-\frac{u} {\lambda_{z_m} \sigma_0^2}\right) \prod\limits_{p \ne m, n} \left(1 - \exp \left(-\frac{u} {\lambda_{z_p} \sigma_0^2} \right)\right) \text{d} u, \nonumber
\end{eqnarray}
where $h_1(\tilde \lambda_{z_n}) = \tilde \lambda_{z_n}^2 \gamma^2 \sigma_\alpha^2 + \tilde \lambda_{z_n} \sigma_0^2$ and $h_2(\tilde \lambda_{z_n}) = h'_1(\tilde \lambda_{z_n}) = 2 \tilde \lambda_{z_n} \gamma^2 \sigma_\alpha^2 + \sigma_0^2$. It is then observed that ${\partial \Gamma_{z_n}} / {\partial \tilde \lambda_{z_n}}$ is larger than zero, and hence we obtain that $\Gamma_{z_n}(\bm \lambda)$ is an increasing function with respect to $\tilde \lambda_{z_n}$ or $\lambda_{z_n}$. By relaxing $\lambda_{z_m}$ to a real variable $\tilde \lambda_{z_m}$ and computing the partial derivative of $\Gamma_{z_n}$ with respect to $\tilde \lambda_{z_m}$, we can similarly verify that $\Gamma_{z_n}(\bm \lambda)$ is a decreasing function of $\lambda_{z_m}$.

Next we prove this theorem. Firstly, when we assume that $\pi_{z_n} \ge \pi_{z_m}$ but $\lambda_{z_n} < \lambda_{z_m}$, we can obtain that
\begin{eqnarray}\label{PF-2}
    \Gamma_{z_n}(\bm \lambda) & = & ~\int\limits_0^\infty \frac{1}{\lambda^2_{z_n} \gamma^2 \sigma_\alpha^2 + \lambda_{z_n} \sigma_0^2} \exp\left(-\frac{u}{\lambda^2_{z_n} \gamma^2 \sigma_\alpha^2 + \lambda_{z_n} \sigma_0^2}\right) \left(1 - \exp\left(-\frac{u}{\lambda_{z_m} \sigma_0^2}\right)\right) G(\bm \lambda_{-z_m, z_n}) \text{d} u \nonumber \\
    & \overset{(a)}{<} & \int\limits_0^\infty \frac{1} {\lambda^2_{z_m} \gamma^2 \sigma_\alpha^2 + \lambda_{z_m} \sigma_0^2} \exp\left(-\frac{u}{\lambda^2_{z_m} \gamma^2 \sigma_\alpha^2 + \lambda_{z_m} \sigma_0^2}\right) \left(1 - \exp\left(-\frac{u}{\lambda_{z_m} \sigma_0^2}\right)\right) G(\bm \lambda_{-z_m, z_n}) \text{d} u \nonumber \\
    & \overset{(b)}{<} & \int\limits_0^\infty \frac{1} {\lambda^2_{z_m} \gamma^2 \sigma_\alpha^2 + \lambda_{z_m} \sigma_0^2} \exp\left(-\frac{u}{\lambda^2_{z_m} \gamma^2 \sigma_\alpha^2 + \lambda_{z_m} \sigma_0^2}\right) \left(1 - \exp\left(-\frac{u}{\lambda_{z_n} \sigma_0^2}\right)\right) G(\bm \lambda_{-z_m, z_n}) \text{d} u \nonumber \\ [5pt]
    & = & \Gamma_{z_m}(\bm \lambda),
\end{eqnarray}
where (a) comes from the fact that $ \Gamma_{z_n}(\bm \lambda)$ is an increasing function with respect to $\lambda_{z_n}$ and (b) is due to $\exp\left(-\frac{u}{\lambda_{z_m} \sigma_0^2}\right) > \exp\left(-\frac{u}{\lambda_{z_n} \sigma_0^2}\right)$. In addition, $\bm \lambda_{-z_m, z_n} \triangleq \{\lambda_{z_1}, \cdots, \lambda_{z_N}\} \big \backslash \{\lambda_{z_m}, \lambda_{z_n}\}$ and $G(\bm \lambda_{-z_m, z_n})$ is expressed as
\begin{equation}\label{PF-3}
    G(\bm \lambda_{-z_m, z_n}) = \prod\limits_{p = 1, p \ne m, n}^N \left(1 - \exp\left(-\frac{u}{\lambda_{z_p} \sigma_0^2}\right) \right).
\end{equation}
It is observed from \eqref{PF-2} that $\Gamma_{z_n} < \Gamma_{z_m}$ when $\lambda_{z_n} < \lambda_{z_m}$. By swapping the values of $\lambda_{z_n}$ and \mbox{$\lambda_{z_m}$ while} keeping $\bm \lambda_{-z_m, z_n}$ unchanged, i.e., $\lambda_{z_n}^{new} = \lambda_{z_m}$, $\lambda_{z_m}^{new} = \lambda_{z_n}$, $\lambda_{z_p}^{new} = \lambda_{z_p}$, $\forall p \ne m, n$, we can see that $\displaystyle \Gamma_{z_n}^{new} = \Gamma_{z_m}$, $\displaystyle \Gamma_{z_m}^{new} = \Gamma_{z_n}$ and $\displaystyle \Gamma_{z_p}^{new} = \Gamma_{z_p}$. The variation of the ASTP after swapping the values of $\lambda_{z_n}$ and $\lambda_{z_m}$ is given by
\begin{equation}\label{PF-4}
    \delta\bar \Gamma_1 = \sum\limits_{p = 1}^N \pi_{z_p} \Gamma_{z_p}^{new} - \sum\limits_{p = 1}^N \pi_{z_p} \Gamma_{z_p} = \Big(\pi_{z_n} - \pi_{z_m}\Big) \Big(\Gamma_{z_m} - \Gamma_{z_n} \Big) > 0,
\end{equation}
which shows that the ASTP $\bar \Gamma_1(\bm \lambda)$ can increase when we exchange the values of $\lambda_{z_n}$ and $\lambda_{z_m}$. By repeating this procedure, we obtain that in order to achieve the maximal $\bar \Gamma_1$, the numbers of used Tx-Rx beam pairs in $\cal B$ should satisfy $\lambda_{s_1} \ge \lambda_{s_2} \ge \cdots \ge \lambda_{s_N}$.

\section{}\label{lemma2-proof}
In order to obtain the lower bound for $\bar \Gamma_1(\bm \lambda)$, we first prove the subsequent inequality:
\begin{equation}\label{B1}
    G_p(\bm \lambda) = \prod\limits_{k = 1}^p \bigg(1 - \exp \bigg(-\frac{u}{\lambda_k \sigma_0^2} \bigg) \bigg) \ge 1 - \sum\limits_{k = 1}^p \exp \bigg(-\frac{u}{\lambda_k \sigma_0^2} \bigg).
\end{equation}
Induction method is used to prove (\ref{B1}). To be specific, when $p = 1$, we can see that (\ref{B1}) is apparently valid. For $p = n - 1$ where $n \ge 2$, we assume that (\ref{B1}) is true. For $p = n$, we have
\begin{eqnarray}\label{B2}
    G_n(\bm \lambda) & = & \prod\limits_{k = 1}^n \bigg(1 -\exp \bigg(-\frac{u}{\lambda_k \sigma_0^2}\bigg)\bigg) = \prod \limits_{k = 1}^{n - 1} \bigg(1 - \exp\bigg( -\frac{u}{\lambda_k \sigma_0^2}\bigg) \bigg) \bigg(1 - \exp\bigg(-\frac{u}{\lambda_n \sigma_0^2} \bigg) \bigg) \nonumber \\
    & \ge & \bigg(1 - \sum\limits_{k = 1}^{n - 1} \exp\bigg(-\frac{u} {\lambda_k \sigma_0^2} \bigg)\bigg)\bigg(1 - \exp\bigg(-\frac{u} {\lambda_n \sigma_0^2} \bigg)\bigg) > 1 - \sum\limits_{k = 1}^n \exp\bigg(-\frac{u}{\lambda_k \sigma_0^2} \bigg),
\end{eqnarray}
which proves (\ref{B1}). By using this inequality, we can rewrite $\Gamma_{s_n}$ as
\begin{eqnarray}\label{B3}
    \Gamma_{s_n} & = & \int\limits_0^\infty \frac{1}{\lambda_{s_n}^2 \gamma^2 \sigma_\alpha^2 + \lambda_{s_n} \sigma_0^2} \exp \left(-\frac{u}{\lambda_{s_n}^2 \gamma^2 \sigma_\alpha^2 + \lambda_{s_n} \sigma_0^2}\right)\prod\limits_{m = 1, m \ne n}^N \left(1 - \exp\left(-\frac{x}{\lambda_{s_m} \sigma_0^2} \right) \right) \text{d} u \nonumber \\
    & \ge & \int\limits_0^\infty \frac{1}{\lambda_{s_n}^2 \gamma^2 \sigma_\alpha^2 + \lambda_{s_n} \sigma_0^2} \exp \left(-\frac{u}{\lambda_{s_n}^2 \gamma^2 \sigma_\alpha^2 + \lambda_{s_n} \sigma_0^2}\right) \left(1 - \sum\limits_{m = 1, m \ne n}^N \exp\left(-\frac{u}{\lambda_{s_m} \sigma_0^2} \right) \right) \text{d} u \nonumber \\
    & = & 1 - \sum\limits_{m = 1, m \ne n}^N \frac{\lambda_{s_m}} {\lambda_{s_n}^2 r_0 + \lambda_{s_n} + \lambda_{s_m}} > 1 - \sum\limits_{m = 1, m \ne n}^N \frac{\lambda_{s_m}} {\lambda_{s_n}^2 r_0 + \lambda_{s_n}} = 1 - \frac{M_{\rm B} - \lambda_{s_n}}{\lambda_{s_n}^2 r_0 + \lambda_{s_n}},
\end{eqnarray}
where $r_0 = \frac{\gamma^2 \sigma_\alpha^2}{\sigma_0^2} = \frac{P N_{\rm T} N_{\rm R} \sigma_\alpha^2} {\sigma_0^2}$, and the ASTP can be therefore lower bounded by
\begin{equation}\label{B4}
    \bar \Gamma_1 (\bm \lambda) = \sum\limits_{n = 1}^N \pi_{s_n} \Gamma_{s_n} > \sum\limits_{n = 1}^N \pi_{s_n} \left(1 - \frac{M_{\rm B} - \lambda_{s_n}}{\lambda_{s_n}^2 r_0 + \lambda_{s_n}} \right) \triangleq \bar \Gamma_1^{\rm lb} (\bm \lambda).
\end{equation}

In order to derive the upper bound of $\bar \Gamma_1 (\bm \lambda)$, we need to prove the subsequent relationship
\begin{equation}\label{B5}
    \sum\limits_{n = 1}^p \frac{(-1)^{n + 1}}{n}\binom{p}{n} = \sum\limits_{n = 1}^p \frac{1}{n}, ~~\forall p \ge 1.
\end{equation}
By exploiting the binomial theorem, we have
\begin{equation}\label{B6}
    \frac{1 - (1 - x)^p}{x} = \sum\limits_{n = 1}^p (-1)^{n + 1} \binom{p}{n} x^{n - 1}, ~~\forall x \in (0, 1].
\end{equation}
Integrating the two sides of (\ref{B6}) from 0 to 1, we can obtain
\begin{equation}\label{B7}
    \int\limits_0^1 \frac{1 - (1 - x)^p}{x} dx = \int\limits_0^1 \sum\limits_{n = 1}^p (-1)^{n + 1} \binom{p}{n} x^{n - 1} d x = \sum\limits_{n = 1}^p \frac{(-1)^{n + 1}}{n} \binom{p}{n}.
\end{equation}
It is worth noting that the right-hand side (RHS) of (\ref{B7}) is the left-hand side (LHS) of (\ref{B5}). For the LHS of (\ref{B7}), we obtain the subsequent equality
\begin{equation}\label{B8}
    \int\limits_0^1 \frac{1 - (1 - x)^p}{x} dx = \int\limits_0^1 \frac{1 - y^p}{1 - y} dy = \int\limits_0^1 (1 +  y + \cdots + y^{p - 1}) d y = \sum\limits_{n = 1}^p \frac{1}{n}.
\end{equation}
By comparing \eqref{B7} and \eqref{B8}, it is seen that (\ref{B5}) has been proven. Next we consider $\Gamma_{s_n}$, which is
\begin{eqnarray}\label{B9}
    \Gamma_{s_n} & = & \int\limits_0^\infty \frac{1}{\lambda_{s_n}^2 \gamma^2 \sigma_\alpha^2 + \lambda_{s_n} \sigma_0^2} \exp \left(-\frac{u}{\lambda_{s_n}^2 \gamma^2 \sigma_\alpha^2 + \lambda_{s_n} \sigma_0^2}\right)\prod\limits_{m = 1, m \ne n}^N \left(1 - \exp\left(-\frac{u}{\lambda_{s_m}\sigma_0^2} \right) \right) \text{d} u \nonumber \\
    & \overset{(a)}{\le} & \int\limits_0^\infty \frac{1}{\lambda_{s_n}^2 \gamma^2 \sigma_\alpha^2 + \lambda_{s_n} \sigma_0^2} \exp \left(-\frac{u}{\lambda_{s_n}^2 \gamma^2 \sigma_\alpha^2 + \lambda_{s_n} \sigma_0^2}\right) \left(1 - \exp\left(-\frac{u} {\sigma_0^2} \right) \right)^{N - 1} \text{d} u \nonumber \\
    & = & 1 - \sum\limits_{m = 1}^{N - 1} \frac{(-1)^{m + 1} \binom{N - 1}{m}}{1 + m (\lambda_{s_n}^2 r_0 + \lambda_{s_n})} \overset{(b)}{\rightarrow} 1 - \sum\limits_{m = 1}^{N - 1} \frac{(-1)^{m + 1} \binom{N - 1}{m}}{m(\lambda_{s_n}^2 r_0 + \lambda_{s_n})} = 1 - \frac{{\cal F}(N - 1)}{\lambda_{s_n}^2 r_0 + \lambda_{s_n}},
\end{eqnarray}
where (a) follows from the fact that $1 - \exp\left(-\frac{u} {\lambda_{s_m}\sigma_0^2}\right)$ is a decreasing function of $\lambda_{s_m}$, and therefore we use $\lambda_{s_m} = 1$ to obtain an upper bound for $\Gamma_{s_n}$, and (b) follows from the fact that $r_0 \gg 1$. By exploiting (\ref{B9}), the ASTP can be upper bounded by
\begin{equation}\label{B10}
    \bar \Gamma_1(\bm \lambda) = \sum\limits_{n = 1}^N \pi_{s_n} \Gamma_{s_n} \le \sum\limits_{n = 1}^N \pi_{s_n} \left[1 - \frac{{\cal F}(N - 1)}{\lambda_{s_n}^2 r_0 + \lambda_{s_n}} \right] \triangleq \bar \Gamma_1^{\rm ub}(\bm \lambda).
\end{equation}

Moreover, we can further approximate $\Gamma_{s_n}$ as
\begin{eqnarray}\label{B12}
    \Gamma_{s_n} & = & \int\limits_0^\infty \frac{1}{\lambda_{s_n}^2 \gamma^2 \sigma_\alpha^2 + \lambda_{s_n} \sigma_0^2} \exp \left(-\frac{u}{\lambda_{s_n}^2 \gamma^2 \sigma_\alpha^2 + \lambda_{s_n} \sigma_0^2}\right)\prod\limits_{m = 1, m \ne n}^N \left(1 - \exp\left(-\frac{u}{\lambda_{s_m}\sigma_0^2} \right) \right) \text{d} u \nonumber \\
    & \approx & \int\limits_0^\infty \frac{1}{\lambda_{s_n}^2 \gamma^2 \sigma_\alpha^2 + \lambda_{s_n} \sigma_0^2} \exp \left(-\frac{u} {\lambda_{s_n}^2 \gamma^2 \sigma_\alpha^2 + \lambda_{s_n} \sigma_0^2} \right) \left(1 - \exp\left(-\frac{u} {\lambda_{\rm avg} \sigma_0^2} \right) \right)^{N - 1} \text{d} u \\
    & = & 1 - \sum\limits_{m = 1}^{N - 1} \frac{(-1)^{m + 1} \binom{N - 1}{m}}{1 + m (\lambda_{s_n}^2 r_0 + \lambda_{s_n}) / \lambda_{\rm avg}} \rightarrow 1 - \sum\limits_{m = 1}^{N - 1} \frac{(-1)^{m + 1} \binom{N - 1}{m}}{m(\lambda_{s_n}^2 r_0 + \lambda_{s_n}) / \lambda_{\rm avg}} = 1 - \frac{\lambda_{\rm avg}{\cal F}(N - 1)}{\lambda_{s_n}^2 r_0 + \lambda_{s_n}}, \nonumber
\end{eqnarray}
where $\lambda_{\rm avg} = \frac{\sum\nolimits_{m \ne n} \lambda_{s_m}} {N - 1} = \frac{M_{\rm B} - \lambda_{s_n}}{N - 1}$. The ASTP is therefore approximated by
\begin{equation}\label{B13}
    \bar \Gamma_1(\bm \lambda) = \sum\limits_{n = 1}^N \pi_{s_n} \Gamma_{s_n} \approx \sum\limits_{n = 1}^N \pi_{s_n} \left[1 - \frac{\lambda_{\rm avg}{\cal F}(N - 1)}{\lambda_{s_n}^2 r_0 + \lambda_{s_n}} \right] \triangleq \bar \Gamma_1^{\rm apx} (\bm \lambda),
\end{equation}
which completes the proof of \textbf{Lemma \ref{lemma2}}.

\section{}\label{lemma3-proof}
To prove that $\bar \Gamma_1^{\rm apx}(\bm \lambda)$ is a concave function with respect to $\tilde \lambda_{s_1}, \cdots, \tilde \lambda_{s_N}$, we compute the first, second and third derivatives of $f(\tilde \lambda_{s_n}) \triangleq (M_{\rm B} - \tilde \lambda_{s_n}) / (\tilde \lambda_{s_n}^2 r_0 + \tilde \lambda_{s_n})$ with respect to $\tilde \lambda_{s_n}$, $\forall n = 1, \cdots, N$, given by
\begin{eqnarray}
    \frac{d f}{d \tilde \lambda_{s_n}} & = & \frac{\tilde \lambda_{s_n}^2 r_0 - 2 M_{\rm B} r_0 \tilde \lambda_{s_n} - M_{\rm B}}{\tilde \lambda_{s_n}^4 r_0^2 + 2 r_0 \tilde \lambda_{s_n}^3 + \tilde \lambda_{s_n}^2}, \label{C1} \\
    \frac{df^2}{d \tilde \lambda_{s_n}^2} & = & \frac{6 M_{\rm B} \tilde \lambda_{s_n}^2 r_0^2 + 6 M_{\rm B} r_0 \tilde \lambda_{s_n} + 2 M_{\rm B} - 2 r_0^2 \tilde \lambda_{s_n}^3} { \tilde \lambda_{s_n}^6 r_0^3 + 3 r_0^2 \tilde \lambda_{s_n}^5 + 3 r_0 \tilde \lambda_{s_n}^4 + \tilde \lambda_{s_n}^3}, \label{C2} \\
    \frac{df^3}{d \tilde \lambda_{s_n}^3} & = & \frac{6 \tilde \lambda_{s_n}^4 r_0^3 - 24 M_{\rm B} r_0^3 \tilde \lambda_{s_n}^3 - 36 M_{\rm B} r_0^2 \tilde \lambda_{s_n}^2 - 24 M_{\rm B} r_0 \tilde \lambda_{s_n} - 6 M_{\rm B}} {\tilde \lambda_{s_n}^8 r_0^4 + 4 r_0^3 \tilde \lambda_{s_n}^7 + 6 r_0^2 \tilde \lambda_{s_n}^6 + 4 r_0 \tilde \lambda_{s_n}^5 + \tilde \lambda_{s_n}^4}.
\end{eqnarray}
Recall that $\tilde \lambda_{s_n} \le M_{\rm B}$, and we can observe that $df^3 / d \tilde \lambda_{s_n}^3 < 0$, which means that $df^2 / d \tilde \lambda_{s_n}^2$ is a decreasing function with respect to $\tilde \lambda_{s_n}$. Due to
\begin{equation}\label{C4}
    \left.\frac{df^2}{d \tilde \lambda_{s_n}^2} \right|_{\tilde \lambda_{s_n} = M_{\rm B}} = \frac{6 M_{\rm B} \tilde \lambda_{s_n}^2 r_0^2 + 6 M_{\rm B} r_0 \tilde \lambda_{s_n} + 2 M_{\rm B} - 2 r_0^2 \tilde \lambda_{s_n}^3}{\tilde \lambda_{s_n}^6 r_0^3 + 3 r_0^2 \tilde \lambda_{s_n}^5 + 3 r_0 \tilde \lambda_{s_n}^4 + \tilde \lambda_{s_n}^3} > 0,
\end{equation}
we can conclude that $f(\tilde \lambda_{s_n})$ is a convex function with respect to $\tilde \lambda_{s_n}$, and therefore $-f(\tilde \lambda_{s_n})$ is a concave function. By noting that $\pi_{s_n} > 0$, $\bar \Gamma_1^{\rm apx}$ is a concave function with respect to $\tilde \lambda_{s_1}, \cdots, \tilde \lambda_{s_N}$, since a nonnegative weighted sum of concave functions is concave \cite{Convex-Opti}. This completes the proof of \textbf{Lemma \ref{lemma3}}.

\section{} \label{lemma4-proof}
Recall that each row of the sensing matrix $\bf A$ can be expressed as
\begin{equation}\label{D-1}
    {\bf a}_{z_m}^{\rm T} = \gamma \alpha \Big(\big[\tilde \nu_{c_m, 1}, \cdots, \tilde \nu_{c_m, X_{\rm T}}\big] \otimes \big[\nu_{a_m, 1}, \cdots, \nu_{a_m, X_{\rm R}}\big]\Big), ~\forall m = 1, \cdots, N.
\end{equation}
In accordance with our previous definition $n = (k \bullet i)_{X_{\rm R}}$, the $n$-th column of $\bf A$ is written as
\begin{equation}
    {\bf A}[:, n] = \gamma \alpha \Big[\underbrace{\nu_{a_1, k} \tilde \nu_{c_1, i}, \cdots, \nu_{a_1, k} \tilde \nu_{c_1, i}}_{\lambda_{z_1}}, \cdots, \underbrace{\nu_{a_N, k} \tilde \nu_{c_N, i}, \cdots, \nu_{a_N, k} \tilde \nu_{c_N, i}}_{\lambda_{z_N}}\Big]^{\rm T},
\end{equation}
and $\bm \xi[n]$ is therefore given by
\begin{eqnarray}\label{D-2}
    \bm \xi[n] & = & {\bf A}^{\rm H}[n, :] {\bf y} = \gamma \alpha^{\ast} \left(\nu_{a_1, k}^{\ast} \tilde \nu_{c_1, i}^{\ast} \sum\limits_{m = 1}^{\lambda_{z_1}} {\bf y}_{a_1, c_1}[m] + \cdots + \nu_{a_N, k}^{\ast} \tilde \nu_{c_N, i}^{\ast} \sum\limits_{m = 1}^{\lambda_{z_N}} {\bf y}_{a_N, c_N}[m] \right) \nonumber \\
    & = & \gamma^2 |\alpha|^2 \sum\limits_{m = 1}^N \bigg(\lambda_{z_m} \nu_{a_m, k}^{\ast} \tilde \nu_{c_m, i}^{\ast} \nu_{a_m, k_1} \tilde \nu_{c_m, i_1} \bigg) + \underbrace{\gamma \alpha^* \sum\limits_{m = 1}^N \nu_{a_m, k}^{\ast} \tilde \nu_{c_m, i}^{\ast} \sum\limits_{m_1 = 1}^{\lambda_{z_m}} n_{p + m_1}}_{\text{noise}},
\end{eqnarray}
where $p = \lambda_{z_1} + \cdots + \lambda_{z_{m - 1}}$. It is worth mentioning that while the ``noise'' terms in $\{\bm \xi[n]\}_{n = 1}^{X}$ are correlated with each other, we ignore their correlations for tractability. The distribution of $\bm \xi[n]$ conditioned on $|\alpha|^2$ is therefore expressed as
\begin{equation}\label{D-3}
    \bm \xi[n] \sim {\cal CN}\left(\gamma^2 |\alpha|^2 \sum\limits_{m = 1}^N \Big[\lambda_{z_m} \nu_{a_m, k}^{\ast} \tilde \nu_{c_m, i}^{\ast} \nu_{a_m, k_1} \tilde \nu_{c_m, i_1} \Big], \gamma^2 |\alpha|^2 \sigma_0^2 \sum\limits_{m = 1}^N \lambda_{z_m} |\nu_{a_m, k} \tilde \nu_{c_m, i}|^2\right).
\end{equation}

Typically, when $n = n_1$ or $k = k_1$ and $i = i_1$, we can write $\bm \xi[n_1]$ as
\begin{equation}\label{D-4}
    \bm \xi[n_1] \sim {\cal CN}\left(\gamma^2 |\alpha|^2 \sum\limits_{m = 1}^N \lambda_{z_m} |\nu_{a_m, n_1} \tilde \nu_{c_m, i_1}|^2, \gamma^2 |\alpha|^2 \sigma_0^2 \sum\limits_{m = 1}^N \lambda_{z_m} |\nu_{a_m, n_1} \tilde \nu_{c_m, i_1}|^2\right).
\end{equation}
By using the union bound, we can rewrite \eqref{NOBP-13} as
\begin{eqnarray}\label{D-5}
    \Gamma_{n_1, |\alpha|^2} & = & 1 - \Pr\Bigg(\bigcup\limits_{n = 1, n \ne n_1}^X \big|\bm \xi[n_1]\big|^2 < \big|\bm \xi[n]\big|^2 ~\Big |~ \alpha \Bigg) \nonumber \\
    & \ge & 1 - \sum\limits_{n = 1, n \ne n_1}^X \Pr\Bigg(\big|\bm \xi[n_1]\big|^2 < \big|\bm \xi[n]\big|^2 ~\Big |~ \alpha \Bigg) \nonumber \\
    & = & 1 - \sum\limits_{n = 1, n \ne n_1}^X \Bigg[ Q_1\left(\sqrt{A_{n, n_1} |\alpha|^2}, \sqrt{B_{n, n_1} |\alpha|^2}\right) - \frac{\sum\nolimits_{m = 1}^N \lambda_{z_m} |\nu_{a_m, k_1} \tilde \nu_{c_m, i_1}|^2}{\sum\nolimits_{m = 1}^N \lambda_{z_m} \Big(|\nu_{a_m, k_1} \tilde \nu_{c_m, i_1}|^2 + |\nu_{a_m, k} \tilde \nu_{c_m, i}|^2\Big)} \nonumber \\
    & \times & \exp\left(-\frac{A_{n, n_1} |\alpha|^2 + B_{n, n_1} |\alpha|^2}{2}\right) I_0 \left(\sqrt{A_{n, n_1} B_{n, n_1}}|\alpha|^2\right)\Bigg],
\end{eqnarray}
where $A_{n, n_1}$ and $B_{n, n_1}$ are respectively given by
\begin{equation}\label{D-6}
    A_{n, n_1} = \frac{2 \gamma^2 \left|\sum\nolimits_{m = 1}^N \lambda_{z_m} \nu_{a_m, k}^{\ast} \tilde \nu_{c_m, i}^{\ast} \nu_{a_m, k_1} \tilde \nu_{c_m, i_1} \right|^2}{\sigma_0^2 \sum\nolimits_{m = 1}^N \lambda_{z_m} \Big(|\nu_{a_m, k_1} \tilde \nu_{c_m, i_1}|^2 + |\nu_{a_m, k} \tilde \nu_{c_m, i}|^2\Big)},
\end{equation}
\begin{equation}
    B_{n, n_1} = \frac{2 \gamma^2 \left(\sum\nolimits_{m = 1}^N \lambda_{z_m}|\nu_{a_m, k_1} \tilde \nu_{c_m, i_1}|^2 \right)^2}{\sigma_0^2 \sum\nolimits_{m = 1}^N \lambda_{z_m} \Big(|\nu_{a_m, k_1} \tilde \nu_{c_m, i_1}|^2 + |\nu_{a_m, k} \tilde \nu_{c_m, i}|^2\Big)}.
\end{equation}

By using the result of \cite{Rician-Integral} and after some mathematical manipulations, we can obtain the integral expression of the Marcum $Q_1$ function over the exponential distribution of $|\alpha|^2$ as
\begin{eqnarray}\label{D-7}
    && \int\limits_0^\infty Q_1\left(\sqrt{A_{n, n_1} |\alpha|^2}, \sqrt{B_{n, n_1} |\alpha|^2}\right) \frac{1}{\sigma_\alpha^2} \exp\left(-\frac{|\alpha|^2}{\sigma_\alpha^2}\right) \text{d} |\alpha|^2
    \nonumber \\
    && ~~~~= \frac{1}{2} - \frac{(B_{n, n_1} - A_{n, n_1}) \sigma_\alpha^2 - 2}{4 \sqrt{1 + (A_{n, n_1} + B_{n, n_1})\sigma_\alpha^2 + \sigma_\alpha^4 (A_{n, n_1} - B_{n, n_1})^2 \big / 4}}.
\end{eqnarray}
Moreover, by exploiting the integral identity \cite{Integral-Table}, we obtain
\begin{align}\label{D-8}
    \int\limits_0^\infty \exp\left(-\frac{A_{n, n_1} |\alpha|^2 + B_{n, n_1} |\alpha|^2}{2}\right) I_0 \left(\sqrt{A_{n, n_1} B_{n, n_1}}|\alpha|^2\right) \frac{1}{\sigma_\alpha^2} \exp\left(-\frac{|\alpha|^2}{\sigma_\alpha^2}\right) \text{d} |\alpha|^2
    \nonumber \\
    = \frac{1}{4 \sqrt{1 + (A_{n, n_1} + B_{n, n_1})\sigma_\alpha^2 + \sigma_\alpha^4 (A_{n, n_1} - B_{n, n_1})^2 \big / 4}}.
\end{align}
Finally, by integrating \eqref{D-5} over the exponential distribution of $|\alpha|^2$, we obtain
\begin{eqnarray}\label{D-9}
    \Gamma_{n_1} & \ge & 1 - \sum\limits_{n = 1, n \ne n_1}^X \left(\frac{1}{2} - \frac{(B_{n, n_1} - A_{n, n_1}) \sigma_\alpha^2 - 2}{4 \sqrt{1 + (A_{n, n_1} + B_{n, n_1})\sigma_\alpha^2 + \sigma_\alpha^4 (A_{n, n_1} - B_{n, n_1})^2 \big / 4}}\right) \nonumber \\
    & + & \sum\limits_{n = 1, n \ne n_1}^X \Bigg(\frac{\sum\nolimits_{m = 1}^N \lambda_{z_m} |\nu_{a_m, k_1} \tilde \nu_{c_m, i_1}|^2}{\sum\nolimits_{m = 1}^N \lambda_{z_m} \Big[|\nu_{a_m, k_1} \tilde \nu_{c_m, i_1}|^2 + |\nu_{a_m, k} \tilde \nu_{c_m, i}|^2\Big]} \nonumber \\
    & \times & ~\frac{1}{\sqrt{1 + (A_{n, n_1} + B_{n, n_1})\sigma_\alpha^2 + \sigma_\alpha^4 (A_{n, n_1} - B_{n, n_1})^2 \big / 4}} \Bigg),
\end{eqnarray}
which completes the proof of \textbf{Lemma \ref{lemma4}}.
\end{appendices}

\end{document}